\documentclass[11pt,a4paper,english,amssymb,nofootinbib,superscriptaddress,showpacs]{revtex4}
\usepackage{lmodern}

\usepackage[T1]{fontenc}
\usepackage[latin9]{inputenc}
\usepackage{color}
\usepackage{babel}
\usepackage{amsmath}
\usepackage{amssymb}
\usepackage{esint}
\usepackage[unicode=true,pdfusetitle,
 bookmarks=true,bookmarksnumbered=false,bookmarksopen=false,
 breaklinks=false,pdfborder={0 0 1},backref=false,colorlinks=false]
 {hyperref}

\makeatletter

\pdfpageheight\paperheight
\pdfpagewidth\paperwidth

\@ifundefined{textcolor}{}
{%
 \definecolor{BLACK}{gray}{0}
 \definecolor{WHITE}{gray}{1}
 \definecolor{RED}{rgb}{1,0,0}
 \definecolor{GREEN}{rgb}{0,1,0}
 \definecolor{BLUE}{rgb}{0,0,1}
 \definecolor{CYAN}{cmyk}{1,0,0,0}
 \definecolor{MAGENTA}{cmyk}{0,1,0,0}
 \definecolor{YELLOW}{cmyk}{0,0,1,0}
 }

\usepackage{lmodern}\usepackage{babel}

\@ifundefined{textcolor}{}{%
 \definecolor{BLACK}{gray}{0}
 \definecolor{WHITE}{gray}{1}
 \definecolor{RED}{rgb}{1,0,0}
 \definecolor{GREEN}{rgb}{0,1,0}
 \definecolor{BLUE}{rgb}{0,0,1}
 \definecolor{CYAN}{cmyk}{1,0,0,0}
 \definecolor{MAGENTA}{cmyk}{0,1,0,0}
 \definecolor{YELLOW}{cmyk}{0,0,1,0}
 }

\@ifundefined{definecolor}{\@ifundefined{definecolor}
 {\@ifundefined{definecolor}
 {\usepackage{color}}{}
}{}
}{}
\usepackage{latexsym}\usepackage{bm}

\usepackage{babel}

\usepackage{babel}

\makeatother

\begin{document}

\title{Some exact solutions of \emph{all }$f\left(R_{\mu\nu}\right)$ theories
in three dimensions}

\author{Metin Gürses}

\email{gurses@fen.bilkent.edu.tr}

\affiliation{{\small Department of Mathematics, Faculty of Sciences}\\
 {\small Bilkent University, 06800 Ankara, Turkey}}

\author{Tahsin Ça\u{g}r\i{} \c{S}i\c{s}man}

\email{tahsin.c.sisman@gmail.com}

\affiliation{Department of Physics,\\
 Middle East Technical University, 06800 Ankara, Turkey}

\author{Bayram Tekin}

\email{btekin@metu.edu.tr}

\affiliation{Department of Physics,\\
 Middle East Technical University, 06800 Ankara, Turkey}

\date{\today}
\begin{abstract}
We find constant scalar curvature Type-N and Type-D solutions in\emph{
all }higher curvature gravity theories with actions of the form $f\left(R_{\mu\nu}\right)$
that are built on the Ricci tensor, but not on its derivatives. In
our construction, these higher derivative theories inherit some of
the previously studied solutions of the cosmological topologically
massive gravity and the new massive gravity field equations, once
the parameters of the theories are adjusted. Besides the generic higher
curvature theory, we have considered in some detail the examples of
the quadratic curvature theory, the cubic curvature theory, and the
Born-Infeld extension of the new massive gravity.
\end{abstract}

\pacs{04.60.Kz, 04.20.Jb, 04.60.Rt, 04.50.Kd}

\maketitle
{\footnotesize \tableofcontents{}}{\footnotesize \par}

\section{Introduction}

In $2+1$ dimensions, various higher curvature modifications of Einstein's
theory, such as the new massive gravity (NMG) \cite{BHT-PRL,BHT-PRD},
a specific cubic curvature gravity \cite{Sinha} and the Born-Infeld
gravity \cite{BINMG} attracted attention recently. NMG provides a
unitary non-linear extension of the Fierz-Pauli mass in both flat
and maximally symmetric constant curvature backgrounds. For anti-de
Sitter (AdS) backgrounds, NMG has a drawback: bulk and boundary unitarity
is in conflict and hence does not fit well into the AdS/CFT picture.
In \cite{Sinha}, a cubic extension of NMG was given which again has
this conflict. A simple, in principle infinite order (in curvature)
extension of NMG in terms of a Born-Infeld gravity, dubbed as BINMG,
was introduced in \cite{BINMG} which again has this bulk-boundary
unitarity conflict. Finally, in \cite{Gullu-AllUni3D}, \emph{all}
bulk and boundary unitary theories in three dimensions were constructed.
These theories should be at least cubic in curvature, if the contractions
of Ricci tensor are used. Linearized excitations in these models have
been studied.

This work is devoted to the study of some exact solutions of all $f\left(R_{\mu\nu}\right)$
theories in three dimensions that include the quadratic, cubic and
BINMG theories as subclasses. For the quadratic and cubic curvature
theories, we will not restrict ourselves to the unitary models but
study the most generic theories. Some exact solutions of NMG were
given in \cite{Clement-Warped,Giribet,Clement-Null_Killing,BHT-PRD,Gurses-Killing,Aliev-PRL,Aliev-PLB,Aliev-PRD}
(and also see \cite{Bakas} for the solutions in the {}``generalized
NMG'' that includes the gravitational Chern-Simons term \cite{DJT-PRL,DJT-Annals}
in addition to the Einstein and the quadratic terms). Save for the
maximally symmetric solutions and AdS-waves \cite{Gullu_Gurses} and
black holes, to the best of our knowledge, a general approach to the
solutions of the general quadratic theory or the more general $f\left(R_{\mu\nu}\right)$
theories has not appeared yet (some Type-III and Type-N solutions
of the $D$-dimensional quadratic gravity were given in \cite{Pravda}).
For cubic curvature theories and for BINMG some solutions were found
before%
\footnote{More recently, Type-N solutions of BINMG and extended NMG theories
appeared in \cite{Aliev-Extended}%
} \cite{Sinha,Nam,Gullu-cfunc,Ghodsi-BH,Ghodsi-AdS-like}. In this
paper we will give a systematic way of finding solutions in these
theories. As it will be clear from the field equations of these theories,
without some symmetry assumption, finding solutions is almost hopeless.
The assumption of the existence of a Killing vector highly restricts
the geometry in three dimensions \cite{Gurses-Killing}, the solutions
of NMG and topologically massive gravity (TMG) under this assumption
include some classes of Type-N and Type-D solutions. However, even
without a symmetry assumption, in addition to the above solutions,
some new Type-N and Type-D solutions of NMG were found in \cite{Aliev-PRL,Aliev-PLB,Aliev-PRD}
using the tetrad formalism. Furthermore, in \cite{Aliev-PRL,Aliev-PLB,Aliev-PRD},
the solutions of NMG inherited from the solutions of TMG are found
by relating the field equations of the two theories. A similar technique
will be applied in this paper to find large classes of solutions of
all $f\left(R_{\mu\nu}\right)$ gravity theories.

The main result of this work is to introduce a technique for finding
exact solutions of any higher curvature gravity theory in three dimensions
from the solutions of TMG and NMG. The technique is based on the observation
that in $D=3$ for constant scalar invariant (CSI) spacetimes%
\footnote{The scalar invariants that we mention are the ones constructed by
contractions of the Ricci tensor but not its derivatives.%
} of Type N and Type D, the field equations of any higher curvature
gravity theory with a generic Lagrangian $f\left(R_{\mu\nu}\right)$
reduce to the field equations of NMG whose form is also same as the
TMG field equations written in the quadratic form. With this fact,
one can obtain new solutions of $f\left(R_{\mu\nu}\right)$ theories
by just relating their parameters to the parameters of TMG and/or
NMG. We have used the results obtained for the $f\left(R_{\mu\nu}\right)$
theory for generic quadratic and cubic curvature gravity theories
in addition to BINMG, and found new solutions for these theories which
are inherited from the Type-N and Type-D solutions of TMG, compiled
in \cite{Chow-Classify}, and NMG found in \cite{Gurses-Killing,Aliev-PRL,Aliev-PLB,Aliev-PRD}.

The layout of the paper is as follows: in Section II, we recapitulate
the algebraic classification of curvature in three dimensions. In
Section III, we derive the quadratic form of the TMG and NMG equations.
Then, we give our main result about $f\left(R_{\mu\nu}\right)$ theories
as a theorem. Section IV is the bulk of the paper where we discuss
the solutions of the $f\left(R_{\mu\nu}\right)$ theories. In Section
V, as an application, we find the solutions of the quadratic and cubic
curvature gravity and BINMG. In appendices, we give some relevant
variations and the field equations of the cubic curvature theory.

\section{Algebraic Classification of Curvature in Three Dimensions\label{sec:Algebraic-Classification}}

In searching for exact solutions, Petrov or Ricci-Segre classification
in three dimensions plays an important role \cite{Wainwright}. Hence,
we briefly review the classification of the exact solutions of TMG
given in \cite{Chow-Classify}. The action of TMG with a cosmological
constant\textcolor{blue}{}%
\footnote{Our signature convention is $\left(-,+,+\right)$ which is opposite
of the original TMG paper; therefore, to account for the {}``wrong''
sign Einstein-Hilbert term, we need to put an overall minus sign to
the action. Note that the sign of $\mu$ is undetermined in TMG. Furthermore,
the overall gravity-matter coupling in the TMG action is taken to
be 1, that is $\kappa=1$, in order to reduce the number of parameters.%
} \cite{DJT-PRL,DJT-Annals} 
\begin{equation}
I=-\int d^{3}x\sqrt{-g}\left[R-2\Lambda+\frac{1}{2\mu}\eta^{\alpha\beta\gamma}\Gamma_{\alpha\nu}^{\mu}\left(\partial_{\beta}\Gamma_{\gamma\mu}^{\nu}+\frac{2}{3}\Gamma_{\beta\rho}^{\nu}\Gamma_{\gamma\mu}^{\rho}\right)\right],
\end{equation}
yields the source-free TMG equations 
\begin{equation}
R_{\mu\nu}-\frac{1}{2}g_{\mu\nu}R+\Lambda g_{\mu\nu}+\frac{1}{\mu}C_{\mu\nu}=0,\label{ctmg}
\end{equation}
where $C_{\mu\nu}$ is the symmetric, traceless and covariantly conserved
Cotton tensor defined as 
\begin{equation}
C_{\mu\nu}\equiv\eta_{\mu\alpha\beta}\nabla^{\alpha}\left(R_{\nu}^{\beta}-\delta_{\nu}^{\beta}\frac{R}{4}\right).\label{cotton}
\end{equation}
Here, the Levi-Civita tensor is given as $\eta_{\mu\sigma\rho}=\sqrt{-g}\varepsilon_{\mu\sigma\rho}$
with $\varepsilon_{012}=+1$ and $g\equiv\mbox{det}\left[g_{\mu\nu}\right]$.
Taking the trace of (\ref{ctmg}) gives $R=6\Lambda$. Therefore,
the Cotton tensor in TMG becomes $C_{\mu\nu}=\eta_{\mu\sigma\beta}\nabla^{\sigma}R_{\nu}^{\beta}$.
Using this, the field equations (\ref{ctmg}) becomes 
\begin{equation}
R_{\mu\nu}-\frac{1}{3}g_{\mu\nu}R+\frac{1}{\mu}\eta_{\mu\alpha\beta}\nabla^{\alpha}R_{\nu}^{\beta}=0,\label{ctmg_without_cotton}
\end{equation}
and defining the traceless Ricci tensor $S_{\mu\nu}\equiv R_{\mu\nu}-\frac{1}{3}g_{\mu\nu}R$
further reduces the field equations to
\begin{equation}
\mu S_{\mu\nu}=-C_{\mu\nu}.\label{eq:ctmg_with_S-C}
\end{equation}

Classification of three dimensional spacetimes can be done either
using the eigenvalues and the eigenvectors of the up-down Cotton tensor
($C_{\nu}^{\mu}$) \cite{Hehl} (in analogy with the four-dimensional
Petrov classification of the Weyl tensor), or the traceless Ricci
tensor ($S_{\nu}^{\mu}$) (in analogy with the Segre classification).
Since $C_{\nu}^{\mu}$ and $S_{\nu}^{\mu}$ are related through (\ref{eq:ctmg_with_S-C}),
for solutions of TMG Segre and Petrov classifications coincide. As
noted in \cite{Chow-Classify}, to determine the eigenvalues of $S_{\mu}^{\nu}$
and their algebraic multiplicities one can compute the two scalar
invariants 
\begin{equation}
I\equiv S_{\mu}^{\nu}S_{\nu}^{\mu},\qquad J\equiv S_{\mu}^{\sigma}S_{\nu}^{\mu}S_{\sigma}^{\nu}.\label{eq:I_J_defns}
\end{equation}
Petrov-Segre Types O, N and III satisfy $I=J=0$; while Types ${\rm D}_{t}$,
${\rm D}_{s}$ and II satisfy $I^{3}=6J^{2}\ne0$. Finally, the most
general types $\text{I}_{R}$ and $\text{I}_{C}$ satisfy $I^{3}>6J^{2}$
and $I^{3}<6J^{2}$ respectively.

For Type-N spacetimes, the canonical form of the traceless Ricci tensor
is
\begin{equation}
S_{\mu\nu}=\rho\,\xi_{\mu}\,\xi_{\nu},\label{eq:Can_Smn_TypeN}
\end{equation}
 where $\xi_{\mu}$ is a null Killing vector and $\rho$ is a scalar
function \cite{Gurses-Killing}. On the other hand, Type-D spacetimes
split into two types that are denoted as ${\rm D}_{t}$ for which
the eigenvector of the traceless Ricci tensor is timelike and ${\rm D}_{s}$
for which the eigenvector is spacelike. For both types, the traceless
Ricci tensor takes the form 
\begin{equation}
S_{\mu\nu}=p\left(g_{\mu\nu}-\frac{3}{\sigma}\xi_{\mu}\xi_{\nu}\right),\label{eq:Can_Smn_TypeD}
\end{equation}
where $p$ is a scalar function and $\xi_{\mu}$ is a timelike or
spacelike vector normalized as $\xi_{\mu}\xi^{\mu}\equiv\sigma=\pm1$.

In this work, we focus on CSI Type-N and Type-D spacetimes. Type-N
spacetimes are CSI if the curvature scalar is constant, while a Type-D
spacetime becomes CSI if $R$ and the scalar function $p$ in (\ref{eq:Can_Smn_TypeD})
are constants.%
\footnote{As we will demonstrate below, the independent scalar invariants of
a three dimensional spacetime are $R$, $S_{\nu}^{\mu}S_{\mu}^{\nu}$,
$S_{\rho}^{\mu}S_{\mu}^{\nu}S_{\nu}^{\rho}$ which satisfy $S_{\nu}^{\mu}S_{\mu}^{\nu}=6p^{2}$,
$S_{\rho}^{\mu}S_{\mu}^{\nu}S_{\nu}^{\rho}=-6p^{3}$ for Type-D spacetimes
requiring constancy of $p$. %
}

\section{A Method to Generate Solutions of $f\left(R_{\mu\nu}\right)$ Theories\label{sec:Method}}

In order to state our main result, first we need to discuss the form
of the field equations of TMG in the quadratic form and the field
equations of NMG for Type-N and Type-D metrics. One can put (\ref{ctmg_without_cotton})
in a second order (wave-like) equation in the Ricci tensor as follows.
Multiplying (\ref{ctmg_without_cotton}) with $\eta_{\phantom{\mu}\sigma\rho}^{\mu}$
and using 
\begin{equation}
\eta^{\mu\alpha\beta}\eta_{\mu\sigma\rho}=-\left(\delta_{\sigma}^{\alpha}\delta_{\rho}^{\beta}-\delta_{\rho}^{\alpha}\delta_{\sigma}^{\beta}\right),\label{convention}
\end{equation}
then, taking the divergence of the resultant equation, one arrives
at the desired equation
\begin{equation}
\square R_{\mu\nu}=\mu^{2}\left(R_{\mu\nu}-2\Lambda g_{\mu\nu}\right)+3R_{\mu\lambda}R_{\nu}^{\lambda}-g_{\mu\nu}R_{\sigma}^{\rho}R_{\rho}^{\sigma}-\frac{3}{2}RR_{\mu\nu}+\frac{1}{2}g_{\mu\nu}R^{2},\label{box_ricci_tensor}
\end{equation}
whose $\Lambda=0$ version was given in \cite{DJT-PRL,DJT-Annals}.
In fact, in the spirit of \cite{DJT-PRL,DJT-Annals}, one can get
the same result with the help of the operator 
\begin{equation}
\mathcal{O}_{\mu\nu}^{\phantom{\mu\nu}\lambda\sigma}\left(\mu\right)\equiv\delta_{\mu}^{\lambda}\delta_{\nu}^{\sigma}-\frac{1}{2}g_{\mu\nu}g^{\lambda\sigma}\left(1-\frac{2\Lambda}{R}\right)+\frac{1}{\mu}\eta_{\mu}^{\phantom{\mu}\alpha\beta}\left(\delta_{\beta}^{\lambda}\delta_{\nu}^{\sigma}-\frac{1}{4}g^{\lambda\sigma}g_{\nu\beta}\right)\nabla_{\alpha},
\end{equation}
 namely, $\mu^{2}\mathcal{O}_{\alpha\beta}^{\phantom{\alpha\beta}\mu\nu}\left(-\mu\right)\mathcal{O}_{\mu\nu}^{\phantom{\mu\nu}\lambda\sigma}\left(\mu\right)R_{\lambda\sigma}=0$
reproduces (\ref{box_ricci_tensor}). It is more transparent to write
the quadratic TMG equation as a pure trace and a traceless part as
\begin{equation}
R=6\Lambda,\label{eq:trace}
\end{equation}
\begin{equation}
\left(\square-\mu^{2}-3\Lambda\right)S_{\mu\nu}=3S_{\mu\rho}S_{\nu}^{\rho}-g_{\mu\nu}S_{\sigma\rho}S^{\sigma\rho},\label{box_tmg}
\end{equation}
where $S_{\mu\nu}$ is the traceless Ricci tensor $S_{\mu\nu}\equiv R_{\mu\nu}-\frac{1}{3}g_{\mu\nu}R$.
It is important to note that every solution of TMG (\ref{ctmg}) solves
(\ref{box_tmg}), but not every solution of the latter solve the former.
For Type-N spacetimes, (\ref{box_tmg}) reduce to 
\begin{equation}
\square S_{\mu\nu}=\left(\mu^{2}+3\Lambda\right)S_{\mu\nu},\label{eq:Trless_TypeN_TMG}
\end{equation}
while for Type-D spacetimes, it becomes
\begin{equation}
\square S_{\mu\nu}=\left(\mu^{2}+3\Lambda-3p\right)S_{\mu\nu}.\label{eq:box_tmg_TypeD}
\end{equation}

Besides using the solutions of TMG, we will also use the solutions
of NMG in order to find solutions to the $f\left(R_{\nu}^{\mu}\right)$
type theories. Hence, let us write the field equations of NMG. In
order to create a parametrization difference between NMG and the generic
quadratic curvature theory that we study, and to directly use the
results given in \cite{Aliev-PRL,Aliev-PLB,Aliev-PRD}, it is better
to prefer the parametrization of NMG used in these works. Then, let
us take the action of NMG as%
\footnote{Notice that we introduce an overall minus sign to the action, \emph{with
the assumption that $G>0$}. Because, in order that NMG defines a
unitary theory, one needs the {}``wrong'' sign Einstein-Hilbert
term.%
}
\begin{equation}
I_{\text{NMG}}=-\frac{1}{16\pi G}\int d^{3}x\sqrt{-g}\,\left[R-2\lambda-\frac{1}{m^{2}}\left(R_{\nu}^{\mu}R_{\mu}^{\nu}-\frac{3}{8}R^{2}\right)\right].
\end{equation}
One can get the field equations of NMG by using the field equations
of generic cubic curvature gravity given in (\ref{pure_trace}) and
(\ref{box_traceless_ricci_ten}) as
\begin{equation}
S_{\mu\nu}S^{\mu\nu}+m^{2}R-\frac{1}{24}R^{2}=6m^{2}\lambda,\label{eq:NMG_trace}
\end{equation}
and
\begin{equation}
\left(\square-m^{2}-\frac{5}{12}R\right)S_{\mu\nu}=4\left(S_{\mu\rho}S_{\nu}^{\rho}-\frac{1}{3}g_{\mu\nu}S_{\sigma\rho}S^{\sigma\rho}\right)+\frac{1}{4}\left(\nabla_{\mu}\nabla_{\nu}-\frac{1}{3}g_{\mu\nu}\square\right)R.\label{eq:NMG_Trless}
\end{equation}
Here, the trace field equation is in the form given in \cite{Aliev-PLB,Aliev-PRD},
while the traceless field equation corresponds to the equation 
\begin{equation}
\left({D\hskip-.25truecm\slash\,}^{2}-m^{2}\right)S_{\mu\nu}=T_{\mu\nu},\label{eq:NMG_Trless_AA-form}
\end{equation}
where the operator $\ensuremath{{D\hskip-.25truecm\slash}\,}$ is
defined through its action on a symmetric tensor $\ensuremath{\Phi_{\mu\nu}}$
as
\begin{equation}
{D\hskip-.25truecm\slash}\,\Phi_{\mu\nu}\equiv\frac{1}{2}\left(\eta_{\mu}^{\phantom{\mu}\alpha\beta}\nabla_{\beta}\Phi_{\nu\alpha}+\eta_{\nu}^{\phantom{\nu}\alpha\beta}\nabla_{\beta}\Phi_{\mu\alpha}\right),
\end{equation}
and

\begin{equation}
T_{\mu\nu}=S_{\mu\rho}S_{\nu}^{\rho}-\frac{1}{3}g_{\mu\nu}S_{\sigma\rho}S^{\sigma\rho}-\frac{R}{12}S_{\mu\nu}.
\end{equation}
It is important to note that two forms of the traceless field equations
of NMG, (\ref{eq:NMG_Trless}) and (\ref{eq:NMG_Trless_AA-form}),
are equivalent whether or not the scalar curvature $R$ is constant.
Now, let us write the field equations of NMG for Type-N spacetimes.
The trace equation (\ref{eq:NMG_trace}) becomes
\begin{equation}
m^{2}R-\frac{1}{24}R^{2}=6m^{2}\lambda,\label{eq:NMG_TypeN_Tr}
\end{equation}
which implies that the scalar curvature is constant. The traceless
field equation (\ref{eq:NMG_Trless}) takes the form
\begin{equation}
\square S_{\mu\nu}=\left(m^{2}+\frac{5}{12}R\right)S_{\mu\nu}.\label{eq:NMG_TypeN_Trless}
\end{equation}
after using the constancy of the scalar curvature. It is easy to see
that the equations (\ref{eq:Trless_TypeN_TMG}) and (\ref{eq:NMG_TypeN_Trless})
are the same equations with different parametrizations of the constant
parts which are related as
\begin{equation}
\mu^{2}=m^{2}-\frac{R}{12}.\label{eq:mu_in_AA_TypeN_soln}
\end{equation}
By using this observation, the Type-N solution of NMG that are based
on the Type-N solution of TMG are found in \cite{Aliev-PRL,Aliev-PLB}.
On the other hand, once the Type-D ansatz is inserted to the NMG equations
(\ref{eq:NMG_trace}) and (\ref{eq:NMG_Trless}), one obtains 
\begin{equation}
6p^{2}+m^{2}R-\frac{1}{24}R^{2}=6m^{2}\lambda,\label{eq:Tr_NMG_Type-D}
\end{equation}
and
\begin{equation}
\left(\square-m^{2}-\frac{5}{12}R+4p\right)S_{\mu\nu}=\frac{1}{4}\left(\nabla_{\mu}\nabla_{\nu}-\frac{1}{3}g_{\mu\nu}\square\right)R.\label{eq:Trless_NMG_Type-D}
\end{equation}
Since we are interested in the constant scalar curvature solutions
of the $f\left(R_{\nu}^{\mu}\right)$ theory, implementing this assumption
in (\ref{eq:Tr_NMG_Type-D}) implies that $p$ is also a constant
and (\ref{eq:Trless_NMG_Type-D}) takes the form 
\begin{equation}
\square S_{\mu\nu}=\left(m^{2}+\frac{5}{12}R-4p\right)S_{\mu\nu}.\label{eq:Trless_NMG_Type-D_const_R}
\end{equation}
As in the Type-N case, (\ref{eq:Trless_NMG_Type-D_const_R}) and the
traceless field equation of TMG for Type-D spacetimes given in (\ref{eq:box_tmg_TypeD})
are the same equation with different parametrizations which are related
by
\begin{equation}
\mu^{2}=m^{2}-\frac{R}{12}-p.\label{eq:mu-m-R-p}
\end{equation}
This observation led to the Type-D solutions of NMG based on the Type-D
solutions of TMG \cite{Gurses-Killing,Aliev-PRL}.

After the above discussion of the field equations of TMG and NMG,
let us focus on the $f\left(R_{\mu\nu}\right)$ theory. As we will
show in the next section, for the CSI Type-N and Type-D spacetimes,
the trace field equation of generic $f\left(R_{\mu\nu}\right)$ theory
determines the constant scalar curvature in terms of the parameters
of the theory; while the traceless field equations reduce to the form
\begin{equation}
\left(\square-c\right)S_{\mu\nu}=0,\label{eq:Wave_eqn_in_Smn}
\end{equation}
where $c$ is a function of the parameters of the theory%
\footnote{For Type-D spacetimes, $c$ also depends on $p$ appearing in (\ref{eq:Can_Smn_TypeD}).%
}. This fact leads us to our main solution inheritance result:

\vspace{0.3cm}

\noindent \textbf{Theorem}: \emph{A Type-N or Type-D solution of TMG
or NMG that has constant scalar curvature generates a solution of
generic $f\left(R_{\mu\nu}\right)$ theory provided that the relations
between the parameters of the corresponding theories, which are obtained
by putting the solution of TMG or NMG as an ansatz into the field
equations of the $f\left(R_{\mu\nu}\right)$ theory, are satisfied.}

\vspace{0.3cm}

\noindent \textbf{Proof}: For Type-N and Type-D spacetimes of constant
scalar curvature, the traceless field equations of TMG, NMG and generic
$f\left(R_{\mu\nu}\right)$ theory take the same wave like equation
for $S_{\mu\nu}$ given in (\ref{eq:Trless_TypeN_TMG}, \ref{eq:box_tmg_TypeD}),
(\ref{eq:NMG_TypeN_Trless}, \ref{eq:Trless_NMG_Type-D}) and (\ref{eq:Wave_eqn_in_Smn}),
respectively. Hence, the main relation between the parameters of the
corresponding theories can be obtained by simply replacing the term
$\square S_{\mu\nu}$ in the traceless field equation of the $f\left(R_{\mu\nu}\right)$
theory by use of the field equations of TMG or NMG. Besides, for Type-D
solutions of TMG (NMG), a specific relation between $\mu^{2}$ ($m^{2}$),
the scalar curvature and $p$ appearing in (\ref{eq:Can_Smn_TypeD})
needs to be satisfied. Finally, the trace field equation of the $f\left(R_{\mu\nu}\right)$
theory determines the scalar curvature in terms of the theory parameters.
Provided that this set of equations relating the parameters in the
corresponding theories is solved, one manages to map the constant
scalar curvature Type-N and Type-D solutions of TMG or NMG to solutions
of the $f\left(R_{\mu\nu}\right)$ theory through these relations.

\vspace{0.3cm}

In the following sections, for the the $f\left(R_{\mu\nu}\right)$
theory, we give the explicit forms of the relations mentioned above
and apply the results to quadratic and cubic curvature theories and
BINMG.

\section{Solutions of $f\left(R_{\mu\nu}\right)$ Theories in Three Dimensions}

In three dimensions, the Riemann tensor does not carry more information
than the Ricci tensor, hence a generic higher curvature theory is
just built on the contractions of the Ricci tensor as
\begin{equation}
I=\int d^{3}x\,\sqrt{-g}\left[\frac{1}{\kappa}\left(R-2\Lambda_{0}\right)+\sum_{n=2}^{\infty}\sum_{\underset{i\ne1}{i=0}}^{n}\sum_{j}a_{ni}^{j}\,\left(R_{\nu}^{\mu}\right)_{j}^{i}\, R^{n-i}\right],\label{eq:Higher_curvature_action}
\end{equation}
where the superscript $i$ in $\left(R_{\nu}^{\mu}\right)_{j}^{i}$
represents the number of Ricci tensors in the term, while the summation
on the subscript $j$ represents the number of possible ways to contract
the $i$ number of Ricci tensors. Each higher curvature combination
has a different coupling denoted by $a_{ni}^{j}$. In the summation
over $i$, the value of 1 is not allowed, simply because it just yields
the scalar curvature upon contraction which is accounted for. For
a given $i$, finding the possible ways of contracting the $i$ number
of Ricci tensors is a counting problem of finding the sequences of
integers satisfying %
\footnote{For the construction of all possible terms at a given order $n$,
see also \cite{Paulos}.%
}
\begin{equation}
i=\sum_{r=1}^{r_{\text{max}}}s_{r};\qquad s_{r}\le s_{r+1},\quad s_{1}\ge2.\label{eq:Ricci_cont_seq}
\end{equation}
Each number in the sequence represents a scalar form involving that
number of Ricci tensors contracted as
\begin{equation}
R_{\mu_{s_{r}}}^{\mu_{1}}\prod_{i=2}^{s_{r}}R_{\mu_{i-1}}^{\mu_{i}}.
\end{equation}
As an example, let us discuss the terms appearing at the curvature
order $n=7$. Even though, the example seems cumbersome, it is quite
useful to understand the counting problem here. The $i$ summation
in (\ref{eq:Higher_curvature_action}) consists of the following $7$
terms;
\[
R^{7},\quad\left(R_{\nu}^{\mu}\right)^{2}R^{5},\quad\left(R_{\nu}^{\mu}\right)^{3}R^{4},\quad\left(R_{\nu}^{\mu}\right)^{4}R^{3},\quad\left(R_{\nu}^{\mu}\right)^{5}R^{2},\quad\left(R_{\nu}^{\mu}\right)^{6}R,\quad\left(R_{\nu}^{\mu}\right)^{7}.
\]
For $i=4,5,6,7$, the possible sequences satisfying (\ref{eq:Ricci_cont_seq})
are 
\begin{align*}
i=4:\quad & \left(2,2\right),\:\left(4\right);\\
i=5:\quad & \left(2,3\right),\:\left(5\right);\\
i=6:\quad & \left(2,2,2\right),\:\left(2,4\right),\:\left(3,3\right),\:\left(6\right);\\
i=7:\quad & \left(2,2,3\right),\:\left(2,5\right),\:\left(3,4\right),\:\left(7\right).
\end{align*}
For example, for $n=i=7$, the terms are
\begin{align*}
 & R_{\mu_{2}}^{\mu_{1}}R_{\mu_{1}}^{\mu_{2}}R_{\mu_{4}}^{\mu_{3}}R_{\mu_{3}}^{\mu_{4}}R_{\mu_{7}}^{\mu_{5}}R_{\mu_{5}}^{\mu_{6}}R_{\mu_{6}}^{\mu_{7}},\quad R_{\mu_{2}}^{\mu_{1}}R_{\mu_{1}}^{\mu_{2}}R_{\mu_{7}}^{\mu_{3}}R_{\mu_{3}}^{\mu_{4}}R_{\mu_{4}}^{\mu_{5}}R_{\mu_{5}}^{\mu_{6}}R_{\mu_{6}}^{\mu_{7}},\\
 & R_{\mu_{3}}^{\mu_{1}}R_{\mu_{1}}^{\mu_{2}}R_{\mu_{2}}^{\mu_{3}}R_{\mu_{7}}^{\mu_{4}}R_{\mu_{4}}^{\mu_{5}}R_{\mu_{5}}^{\mu_{6}}R_{\mu_{6}}^{\mu_{7}},\quad R_{\mu_{7}}^{\mu_{1}}R_{\mu_{1}}^{\mu_{2}}R_{\mu_{2}}^{\mu_{3}}R_{\mu_{3}}^{\mu_{4}}R_{\mu_{4}}^{\mu_{5}}R_{\mu_{5}}^{\mu_{6}}R_{\mu_{6}}^{\mu_{7}}.
\end{align*}
What is important to realize is that at each order $n$, there is
only one term which cannot be constructed as a multiplication of the
terms that already appear at the lower orders compared to $n$. This
term is $\left(R_{\nu}^{\mu}\right)^{n}$ with the contraction sequence
$\left(n\right)$ that is
\begin{equation}
R_{\mu_{n}}^{\mu_{1}}R_{\mu_{1}}^{\mu_{2}}R_{\mu_{2}}^{\mu_{3}}\dots R_{\mu_{n-1}}^{\mu_{n}}.\label{eq:Ricci_(n)}
\end{equation}
However, it is shown in \cite{Paulos} that for $n>3$, the term (\ref{eq:Ricci_(n)})
can be written as a sum of the other terms appearing in order $n$
by use of the Schouten identities;
\begin{equation}
\delta_{\nu_{1}\nu_{2}\dots\nu_{n}}^{\mu_{1}\mu_{2}\dots\mu_{n}}R_{\mu_{1}}^{\nu_{1}}R_{\mu_{2}}^{\nu_{2}}\dots R_{\mu_{n}}^{\nu_{n}}=0,\quad n>3,\label{eq:Schouten}
\end{equation}
where $\delta_{\nu_{1}\nu_{2}\dots\nu_{n}}^{\mu_{1}\mu_{2}\dots\mu_{n}}$
is the generalized Kronecker delta with the definition
\begin{equation}
\delta_{\nu_{1}\dots\nu_{2n}}^{\mu_{1}\dots\mu_{2n}}\equiv\det\left|\begin{array}{ccc}
\delta_{\nu_{1}}^{\mu_{1}} & \dots & \delta_{\nu_{1}}^{\mu_{2n}}\\
\vdots & \ddots & \vdots\\
\delta_{\nu_{2n}}^{\mu_{1}} & \dots & \delta_{\nu_{2n}}^{\mu_{2n}}
\end{array}\right|.
\end{equation}
The basis for the Schouten identities is the simple fact that in three
dimensions a totally antisymmetric tensor having a rank higher than
3 is identically zero. Therefore, for $n>3$, the new term appearing
at order $n$ given in (\ref{eq:Ricci_(n)}) can be written as a sum
of the terms which involve $n$ curvature forms and are multiplications
of the terms that already appear at the lower orders. This fact implies
that the terms $R$, $R_{\nu}^{\mu}R_{\mu}^{\nu}$, $R_{\rho}^{\mu}R_{\mu}^{\nu}R_{\nu}^{\rho}$
are the only independent curvature combinations, and every other term
that can be constructed by any kind of contraction of any number of
Ricci tensors can be obtained as a function of these three terms \cite{Paulos}.
Therefore, a higher curvature gravity action of the form $f\left(R_{\nu}^{\mu}\right)$
either given in a series expansion or in a closed form can be put
in a form $f\left(R_{\nu}^{\mu}\right)=F\left(R,R_{\nu}^{\mu}R_{\mu}^{\nu},R_{\rho}^{\mu}R_{\mu}^{\nu}R_{\nu}^{\rho}\right)$.

As revealed in the previous sections, working with the traceless Ricci
tensor instead of the Ricci tensor at the equation of motion level
simplifies the computations. Let us consider this change at the action
level, and obtain the field equations in terms of the traceless Ricci
tensor directly. With this change every observation made in the previous
paragraphs remains as the same with one simplification: the Schouten
identity written in terms of the traceless Ricci tensor
\begin{equation}
0=\delta_{\nu_{1}\nu_{2}\dots\nu_{n}}^{\mu_{1}\mu_{2}\dots\mu_{n}}S_{\mu_{1}}^{\nu_{1}}S_{\mu_{2}}^{\nu_{2}}\dots S_{\mu_{n}}^{\nu_{n}},\quad n>3,
\end{equation}
 involve less number of terms than (\ref{eq:Schouten}) due to vanishing
trace of $S_{\mu\nu}$.

Now, let us study the field equation of higher curvature gravity theories.
With the hindsight gained in the previous paragraphs, it is sufficient
and convenient to study the action 
\begin{equation}
I=\int d^{3}x\,\sqrt{-g}F\left(R,A,B\right),\label{eq:Higher_curv_act_in_RAB}
\end{equation}
where
\[
A\equiv S_{\nu}^{\mu}S_{\mu}^{\nu},\qquad B\equiv S_{\rho}^{\mu}S_{\mu}^{\nu}S_{\nu}^{\rho},
\]
and $F$ is either a power series expansion in $\left(R,A,B\right)$,
or an analytic function in these variables. It is worth restating
the arguments on this choice: studying this action is sufficient since
any higher curvature action which involves just the scalars constructed
from only $R_{\nu}^{\mu}$, and not its derivatives, can be put in
this form; on the other hand, it is convenient to study a generic
form of a higher curvature action in $\left(R,A,B\right)$ because
we aim to figure out the general structure of the Type-N and Type-D
solutions whose analysis becomes easy by using the canonical form
of the traceless Ricci tensor. The variation of the action (\ref{eq:Higher_curv_act_in_RAB})
has the form
\begin{equation}
\delta I=\int d^{3}x\,\sqrt{-g}\left(F_{R}\delta R+F_{A}\delta A+F_{B}\delta B-\frac{1}{2}g_{\mu\nu}F\delta g^{\mu\nu}\right),
\end{equation}
where $F_{R}\equiv\frac{\partial F}{\partial R}$, and $F_{A}$, $F_{B}$
are defined similarly. Using the $\delta R$, $\delta A$ and $\delta B$
results given in the appendix, the field equations for the action
(\ref{eq:Higher_curv_act_in_RAB}) become 
\begin{align}
-\frac{1}{2}g_{\mu\nu}F+2F_{A}S_{\mu}^{\rho}S_{\rho\nu}+3F_{B}S_{\mu}^{\rho}S_{\rho\sigma}S_{\nu}^{\sigma}+\left(\Box+\frac{2}{3}R\right)\left(F_{A}S_{\mu\nu}+\frac{3}{2}F_{B}S_{\mu}^{\rho}S_{\rho\nu}\right)\nonumber \\
+\left(g_{\mu\nu}\Box-\nabla_{\mu}\nabla_{\nu}+S_{\mu\nu}+\frac{1}{3}g_{\mu\nu}R\right)\left(F_{R}-F_{B}S_{\sigma}^{\rho}S_{\rho}^{\sigma}\right)\label{eq:Higher_curv_eom}\\
-2\nabla_{\alpha}\nabla_{(\mu}\left(S_{\nu)}^{\alpha}F_{A}+\frac{3}{2}S_{\nu)}^{\rho}S_{\rho}^{\alpha}F_{B}\right)+g_{\mu\nu}\nabla_{\alpha}\nabla_{\beta}\left(F_{A}S^{\alpha\beta}+\frac{3}{2}F_{B}S^{\alpha\rho}S_{\rho}^{\beta}\right) & =0.\nonumber 
\end{align}
A simple observation is that the Type-O spacetimes (for which $S_{\mu\nu}=0$)
with constant scalar curvature satisfy the field equation
\begin{equation}
\frac{3}{2}F-RF_{R}=0.
\end{equation}
Furthermore, note that $A$ and $B$ are zero for Type-N spacetimes;
and they are proportional to $p^{2}$ and $p^{3}$, respectively,
for Type-D spacetimes. Therefore, $F$, $F_{R}$, $F_{A}$, $F_{B}$
are functions of $R$ for Type-N spacetimes; while they are functions
of $R$ and $p$ for Type-D spacetimes. 

Now, let us study the Type-N and Type-D solutions of the $f\left(R_{\nu}^{\mu}\right)$
gravity which are also solutions of the cosmological TMG or NMG. In
finding these solutions, we will assume that the spacetime is CSI.
This assumption implies that the scalar curvature is constant in addition
to the constancy of $p$ for Type-D spacetimes. Without such an assumption,
one cannot proceed unless the explicit form of $F$, that is the action,
is given.

\subsection{Type-N Solutions\label{sec:Type-N_f(Ricci)}}

Recall that for Type-N spacetimes contractions of two and more traceless
Ricci tensors vanish; therefore, for such spacetimes (\ref{eq:Higher_curv_eom})
becomes
\begin{align}
-\frac{1}{2}g_{\mu\nu}F+\left(\Box+\frac{2}{3}R\right)\left(F_{A}S_{\mu\nu}\right)+\left(g_{\mu\nu}\Box-\nabla_{\mu}\nabla_{\nu}+S_{\mu\nu}+\frac{1}{3}g_{\mu\nu}R\right)F_{R}\nonumber \\
-2\nabla_{\alpha}\nabla_{(\mu}\left(S_{\nu)}^{\alpha}F_{A}\right)+g_{\mu\nu}\nabla_{\alpha}\nabla_{\beta}\left(F_{A}S^{\alpha\beta}\right) & =0.\label{eq:TypeN_high_curv_eom}
\end{align}
Constancy of the scalar curvature $R$ implies that $F$, $F_{R}$,
$F_{A}$, $F_{B}$ are all constants, since they only depend on $R$
for Type-N spacetimes. Besides, one has the Bianchi identity $\nabla_{\mu}S_{\nu}^{\mu}=\frac{1}{6}\nabla_{\nu}R=0$
which further simplifies (\ref{eq:TypeN_high_curv_eom}) to 
\begin{equation}
\left(\frac{1}{3}RF_{R}-\frac{1}{2}F\right)g_{\mu\nu}+\left(F_{A}\Box-\frac{1}{3}RF_{A}+F_{R}\right)S_{\mu\nu}=0.
\end{equation}
Since $S_{\mu\nu}$ is traceless, the field equations split into two
parts as
\begin{equation}
\frac{3}{2}F-RF_{R}=0,\label{eq:TypeN_Tr_field_eqn}
\end{equation}
\begin{equation}
\left(F_{A}\Box-\frac{1}{3}RF_{A}+F_{R}\right)S_{\mu\nu}=0,\label{eq:TypeN_Trless_field_eqn}
\end{equation}
which are the trace and the traceless field equations of the higher
curvature gravity theory for Type-N spacetimes with constant curvature.
The first equation determines the scalar curvature, and the second
equation is of the form $\left(\Box-c\left(R;a_{ni}^{j}\right)\right)S_{\mu\nu}=0$,
where $c\left(R;a_{ni}^{j}\right)$ is a constant depending on $R$
and the parameters of the $f\left(R_{\nu}^{\mu}\right)$ theory and
does not vanish generically. Even though we have reduced the complicated
field equations of the generic $f\left(R_{\nu}^{\mu}\right)$ theory
to a Klein-Gordon type equation for $S_{\mu\nu}$, it is still a highly
complicated nonlinear equation for the metric and without further
assumptions such as the existence of symmetries it would be hard to
find explicit solutions. But, the state of affairs is not that bleak,
as we will lay out below, the field equations of TMG (in the quadratic
form) and NMG also reduce to Klein-Gordon type equations for $S_{\mu\nu}$
for Type-N spacetimes%
\footnote{This is the observation which yields the Type-N solutions of NMG which
are also solutions of TMG found in \cite{Gurses-Killing,Aliev-PRL}.%
}. Such solutions in these theories have been studied before.\textcolor{red}{{}
}In \cite{Aliev-PLB}, the Type-N solutions for the the equation $\left(\Box-c\left(R;\mu^{2}\right)\right)S_{\mu\nu}=0$
with constant curvature is studied where the form of $c\left(R;\mu^{2}\right)$
is specifically the one corresponding to the quadratic form of the
TMG field equations.

Now, let us discuss the solutions based on the solutions of TMG and
NMG, separately; and elaborate on the relation between them.

\subsubsection{Solutions based on TMG\label{sub:Solutions_on_TypeN_TMG}}

The Type-N solutions of TMG satisfy the field equations (\ref{eq:trace})
and (\ref{eq:Trless_TypeN_TMG}). Then, requiring that the Type-N
solutions of $f\left(R_{\nu}^{\mu}\right)$ gravity are also solutions
of TMG, the field equations (\ref{eq:TypeN_Tr_field_eqn}) and (\ref{eq:TypeN_Trless_field_eqn})
take the forms
\begin{equation}
F-4\Lambda F_{R}=0,\label{eq:R_eqn_TypeN}
\end{equation}
and
\begin{equation}
\mu^{2}=-\left(\frac{F_{R}}{F_{A}}+\Lambda\right).\label{eq:mu_eqn_TypeN}
\end{equation}
Generically, (\ref{eq:R_eqn_TypeN}) is not an algebraic equation.
If it is solved for the unknown $\Lambda$, its solution together
with (\ref{eq:mu_eqn_TypeN}) fixes $\Lambda$ and $\mu^{2}$ in terms
of the parameters of the higher curvature theory. Once these two equations
are satisfied Type-N solutions of TMG compiled in \cite{Chow-Classify,Chow-Kundt}
also solve the $f\left(R_{\nu}^{\mu}\right)$ theory.

\subsubsection{Solutions based on NMG}

The field equations of NMG for Type-N spacetimes reduce to (\ref{eq:NMG_TypeN_Tr})
and (\ref{eq:NMG_TypeN_Trless}). One should recall that the traceless
field equations of TMG, (\ref{eq:Trless_TypeN_TMG}), and NMG, (\ref{eq:NMG_TypeN_Trless}),
are the same equations with different parametrizations related as
(\ref{eq:mu_in_AA_TypeN_soln}). In \cite{Gurses-Killing,Aliev-PLB},
the Type-N solutions of NMG are parametrized by $R$ and $\mu$ %
\footnote{In fact, (\ref{eq:Trless_TypeN_TMG}) is the equation solved in \cite{Aliev-PLB}
with the definition $\nu^{2}\equiv-R/6$.%
}. Therefore, in order to find the Type-N solutions of the $f\left(R_{\nu}^{\mu}\right)$
theory which are also solutions of NMG, (\ref{eq:R_eqn_TypeN}) and
(\ref{eq:mu_eqn_TypeN}) are again the equations which need to be
satisfied. The Type-N solutions of TMG that we used in the previous
section satisfy the traceless field equation of TMG 
\begin{equation}
\eta_{\mu\alpha\beta}\nabla^{\alpha}S_{\nu}^{\beta}+\mu S_{\mu\nu}=0,
\end{equation}
besides the second order equation (\ref{eq:Trless_TypeN_TMG}). The
general Type-N solution of NMG given in \cite{Aliev-PLB} includes
the TMG based solutions as special limits in addition to the solutions
which only solve the quadratic equation (\ref{eq:Trless_TypeN_TMG}).
Thus, once the Type-N solutions of the $f\left(R_{\nu}^{\mu}\right)$
theory with NMG origin are found, the Type-N solutions based on TMG
are also obtained by considering the limits given in \cite{Aliev-PLB}.
Note that as shown in \cite{Aliev-PRL,Aliev-PLB}, there are two classes
of Type-N solutions of NMG depending on whether the eigenvector $\xi_{\nu}=\partial_{v}$
of $S_{\mu\nu}$ is a Killing vector or not. Here, we have covered
both of these Type-N solutions.

\subsection{Type-D Solutions\label{sec:Type-D-f(Ricci)}}

First of all, we just employ the canonical form of the traceless Ricci
tensor for Type-D spacetimes given in (\ref{eq:Can_Smn_TypeD}) in
the field equations of the $f\left(R_{\nu}^{\mu}\right)$ theory given
in (\ref{eq:Higher_curv_eom}). In the equations, there are rank $\left(0,2\right)$
tensors formed by the contractions of two and three traceless Ricci
tensors. With the use of (\ref{eq:Can_Smn_TypeD}), one can show that
these forms are just linear combinations of the metric and $S_{\mu\nu}$
as
\begin{equation}
S_{\mu}^{\rho}S_{\rho\nu}=p\left(2pg_{\mu\nu}-S_{\mu\nu}\right),\qquad S_{\mu}^{\rho}S_{\rho\sigma}S_{\nu}^{\sigma}=p^{2}\left(3S_{\mu\nu}-2pg_{\mu\nu}\right).\label{eq:2-3_Smn_in_TypeD}
\end{equation}
Therefore, for Type-D spacetimes, the rank $\left(0,2\right)$ tensors
that should appear in the equations of motion of the $f\left(R_{\nu}^{\mu}\right)$
theory are just the metric and the traceless Ricci tensor; and consequently,
(\ref{eq:Higher_curv_eom}) takes the form
\begin{align}
\left(-\frac{1}{2}F+4p^{2}F_{A}-6p^{3}F_{B}\right)g_{\mu\nu}+\left(-2pF_{A}+9p^{2}F_{B}\right)S_{\mu\nu}\nonumber \\
+\left(\Box+\frac{2}{3}R\right)\left[3p^{2}F_{B}g_{\mu\nu}+\left(F_{A}-\frac{3}{2}pF_{B}\right)S_{\mu\nu}\right]\nonumber \\
+\left(g_{\mu\nu}\Box-\nabla_{\mu}\nabla_{\nu}+S_{\mu\nu}+\frac{1}{3}g_{\mu\nu}R\right)\left(F_{R}-6p^{2}F_{B}\right)\label{eq:TypeD_high_curv_eom}\\
-2\nabla_{\alpha}\nabla_{(\mu}\left[\delta_{\nu)}^{\alpha}3p^{2}F_{B}+S_{\nu)}^{\alpha}\left(F_{A}-\frac{3}{2}pF_{B}\right)\right]\nonumber \\
+g_{\mu\nu}\nabla_{\alpha}\nabla_{\beta}\left[3p^{2}F_{B}g^{\alpha\beta}+\left(F_{A}-\frac{3}{2}pF_{B}\right)S^{\alpha\beta}\right] & =0.\nonumber 
\end{align}
If $R$ and $p$ are assumed to be constant (note that $p$ should
be a constant for TMG, and the constancy of $R$ implies the constancy
of $p$ for the NMG case), then $F$, $F_{R}$, $F_{A}$, $F_{B}$
should also be constant due to the fact that they just depend on $R$
and $p$. Then, (\ref{eq:TypeD_high_curv_eom}) reduces to
\begin{equation}
0=\left[-\frac{1}{2}F+\frac{1}{3}RF_{R}+4p^{2}\left(F_{A}-\frac{3}{2}pF_{B}\right)\right]g_{\mu\nu}+\left[F_{R}+\left(F_{A}-\frac{3}{2}pF_{B}\right)\left(\Box-\frac{1}{3}R+4p\right)\right]S_{\mu\nu},
\end{equation}
where the Bianchi identity was also used. Since $S_{\mu\nu}$ is traceless,
the field equations split into two parts as
\begin{equation}
\frac{3}{2}F-RF_{R}-6p^{2}\left(2F_{A}-3pF_{B}\right)=0,\label{eq:TypeD_Tr_field_eqn}
\end{equation}
\begin{equation}
\left[F_{R}+\left(F_{A}-\frac{3}{2}pF_{B}\right)\left(\Box-\frac{1}{3}R+4p\right)\right]S_{\mu\nu}=0.\label{eq:TypeD_Trless_field_eqn}
\end{equation}
They are the trace and the traceless field equations of the $f\left(R_{\nu}^{\mu}\right)$
theory for the Type-D spacetimes with constant $R$ and $p$. In (\ref{eq:TypeD_Tr_field_eqn}),
the unknowns are $p$ and $R$; therefore, unlike the case of the
Type-N spacetimes, the trace field equation does not yield a solution
for $R$. On the other hand, as in the case of Type-N spacetimes,
the only operator in the traceless field equation is the d'Alembertian,
$\square$, so the equation is in the form 
\begin{equation}
\left[\square-c\left(R,p;a_{ni}^{j}\right)\right]S_{\mu\nu}=0,
\end{equation}
where $c\left(R,p;a_{ni}^{j}\right)$ is a constant. For Type-D spacetimes
of constant curvature, the traceless field equations of TMG (in the
quadratic form) and NMG are also in the same Klein-Gordon form where
the functional dependence of $c$ is $c\left(R,p;\mu^{2}\right)$
in the TMG case and $c\left(R,p;m^{2}\right)$ in the NMG case. However,
the set $\left(R,p;\mu^{2}\right)$ of the TMG case and the set $\left(R,p;m^{2}\right)$
of the NMG case are not independent, but satisfy specific algebraic
relations in order to yield Type-D solutions. We use these algebraic
relations to put the traceless field equation of TMG or NMG in the
form $\left[\square-c\left(R,p\right)\right]S_{\mu\nu}=0$. Then,
assuming that the Type-D solutions of either TMG or NMG are also Type-D
solutions of the $f\left(R_{\nu}^{\mu}\right)$ theory, one can replace
$\square S_{\mu\nu}$ with the term $c\left(R,p\right)S_{\mu\nu}$.
The resulting equation together with (\ref{eq:TypeD_Tr_field_eqn})
constitute a coupled set of equations that needs to be satisfied in
order to have constant curvature Type-D solutions of the $f\left(R_{\nu}^{\mu}\right)$
theory which are also solutions of TMG or NMG. Now, let us discuss
the Type-D solutions based on TMG and NMG, and their relation.

\subsubsection{Solutions based on TMG}

For Type-D solutions of the cosmological TMG, as shown in \cite{Nutku,Gurses},
the function $p$ appearing in (\ref{eq:Can_Smn_TypeD}) is a constant
and the vector $\xi^{\mu}$ should be a Killing vector satisfying
\begin{equation}
\nabla_{\mu}\xi_{\nu}=\frac{\mu}{3}\eta_{\mu\nu\rho}\xi^{\rho},
\end{equation}
where, in order to have a solution, $\mu$ should be related to $p$
and $\Lambda$ as
\begin{equation}
\mu^{2}=9\left(p-\Lambda\right).\label{eq:mu-R-p}
\end{equation}
Type-D solutions of TMG automatically satisfy the traceless field
equation of TMG in the quadratic form given in (\ref{eq:box_tmg_TypeD})
which becomes
\begin{equation}
\left[\square+6\left(\Lambda-p\right)\right]S_{\mu\nu}=0,\label{eq:Trless_TypeD_TMG}
\end{equation}
with the help of (\ref{eq:mu-R-p}). If one requires that the Type-D
solutions of the $f\left(R_{\nu}^{\mu}\right)$ theory are also solutions
of TMG, then by using (\ref{eq:Trless_TypeD_TMG}) and the trace field
equation of TMG, that is $R=6\Lambda$, in the field equations of
the $f\left(R_{\nu}^{\mu}\right)$ theory for Type-D spacetimes given
in (\ref{eq:TypeD_Tr_field_eqn}) and (\ref{eq:TypeD_Trless_field_eqn}),
one arrives at
\begin{equation}
F-4\Lambda F_{R}-4p^{2}\left(2F_{A}-3pF_{B}\right)=0,\label{eq:fRmn_TypeD_Eqn1_TMG}
\end{equation}
and
\begin{equation}
F_{R}-\left(2F_{A}-3pF_{B}\right)\left(4\Lambda-5p\right)=0.\label{eq:fRmn_TypeD_Eqn2_TMG}
\end{equation}
This coupled set of equations determines $p$ and $\Lambda$ in terms
of the parameters of the $f\left(R_{\nu}^{\mu}\right)$ theory. One
should keep in mind that the solutions of this set should satisfy
(\ref{eq:mu-R-p}); or, in other words, they determine $\mu^{2}$.
Once $\Lambda$ and $\mu^{2}$ are found, their use in the Type-D
solutions of TMG compiled in \cite{Chow-Classify} which are parametrized
by $\Lambda$ and $\mu^{2}$ yields the Type-D solutions of the $f\left(R_{\nu}^{\mu}\right)$
theory.

\subsubsection{Solutions based on NMG}

As we discussed in Sec.~\ref{sec:Method}, for constant scalar curvature
Type-D spacetimes, the field equations of NMG take the forms (\ref{eq:Tr_NMG_Type-D})
and (\ref{eq:Trless_NMG_Type-D_const_R}), and the latter equation
is nothing but the reparametrized version of the traceless field equation
of TMG for Type-D spacetimes given in (\ref{eq:box_tmg_TypeD}). The
constancy of $R$ implies the constancy of $p$ via (\ref{eq:Tr_NMG_Type-D})
which in turn indicates that the constant scalar curvature solutions
of NMG are CSI spacetimes.

Like the Type-D solutions of TMG, (\ref{eq:Trless_NMG_Type-D_const_R})
is solved, if and only if the parameters $p$, $R$ and $m^{2}$ are
related in a specific way. The equations (\ref{eq:mu-R-p}) and (\ref{eq:mu-m-R-p})
yield 
\begin{equation}
p=\frac{m^{2}}{10}+\frac{17}{120}R.\label{eq:p-R-m2_reln_TMG}
\end{equation}
If (\ref{eq:p-R-m2_reln_TMG}) is satisfied, then there are Type-D
solutions of NMG which are also solutions of TMG with the parameters
given below by (\ref{eq:Type-D_m2}) and (\ref{eq:Type-D_L}) after
using $\alpha=-3\beta/8$, $m^{2}=-\frac{1}{\kappa\beta}$ \cite{Aliev-PRD}.
Furthermore, there are also Type-D solutions which are exclusively
solutions of NMG, but not solutions of TMG. These solutions are separated
into two classes differing with respect to the relations satisfied
by the parameters of NMG and its Type-D solution. This follows whether
$\xi_{\mu}$ is a hypersurface orthogonal Killing-vector or a covariantly
divergence-free vector, not a Killing vector \cite{Aliev-PRD}. For
the covariantly divergence-free vector case, the parameters should
be related as 
\begin{equation}
p=-\frac{R}{3}=-\frac{4}{15}m^{2},\qquad\lambda=\frac{m^{2}}{5},\label{eq:p-R-m2_reln_nonTMG1}
\end{equation}
while for the case of hypersurface orthogonal Killing-vector, the
parameters are related as
\begin{equation}
p=\frac{R}{6}=\frac{2}{3}m^{2},\qquad\lambda=m^{2}.\label{eq:p-R-m2_reln_nonTMG2}
\end{equation}
As it can be observed from (\ref{eq:p-R-m2_reln_nonTMG1}) and (\ref{eq:p-R-m2_reln_nonTMG2}),
these Type-D solutions provided in \cite{Aliev-PRD} are uniquely
parametrized by $m^{2}$. 

By requiring that the Type-D solutions of the $f\left(R_{\nu}^{\mu}\right)$
theory to be also solutions of NMG, one can use (\ref{eq:Trless_NMG_Type-D_const_R})
to reduce the traceless field equation of the $f\left(R_{\nu}^{\mu}\right)$
theory given in (\ref{eq:TypeD_Trless_field_eqn}) to 
\begin{equation}
F_{R}+\left(F_{A}-\frac{3}{2}pF_{B}\right)\left(m^{2}+\frac{1}{12}R\right)=0.\label{eq:fRmn_TypeD_Eqn2_NMG}
\end{equation}
With (\ref{eq:TypeD_Tr_field_eqn}), (\ref{eq:fRmn_TypeD_Eqn2_NMG})
constitute the equations that should be solved in order to write the
parameters of the Type-D solutions of NMG in terms of the parameters
of the $f\left(R_{\nu}^{\mu}\right)$ theory. Besides, the parameters
$p$, $R$ and $m^{2}$ appearing in (\ref{eq:TypeD_Tr_field_eqn})
and (\ref{eq:fRmn_TypeD_Eqn2_NMG}) need to satisfy either one of
the equations (\ref{eq:p-R-m2_reln_TMG}), (\ref{eq:p-R-m2_reln_nonTMG1})
or (\ref{eq:p-R-m2_reln_nonTMG2}). If one chooses to eliminate $m^{2}$
in favor of $p$ and $R$ by using (\ref{eq:p-R-m2_reln_TMG}), then
(\ref{eq:TypeD_Tr_field_eqn}) and (\ref{eq:fRmn_TypeD_Eqn2_NMG})
reduce to (\ref{eq:fRmn_TypeD_Eqn1_TMG}) and (\ref{eq:fRmn_TypeD_Eqn2_TMG})
of the TMG case. Therefore, one obtains the same Type-D solutions
discussed in the previous section, if $p$, $R$ and $m^{2}$ satisfy
(\ref{eq:p-R-m2_reln_TMG}). On the other hand, if these parameters
satisfy (\ref{eq:p-R-m2_reln_nonTMG1}) or (\ref{eq:p-R-m2_reln_nonTMG2}),
then (\ref{eq:TypeD_Tr_field_eqn}) and (\ref{eq:fRmn_TypeD_Eqn2_NMG})
reduce either to
\begin{equation}
F=0,\qquad F_{R}+\frac{2}{3}R\left(2F_{A}+RF_{B}\right)=0,\label{eq:nonTMG1}
\end{equation}
or to 
\begin{equation}
F=0,\qquad F_{R}+\frac{R}{3}\left(F_{A}-\frac{1}{4}RF_{B}\right)=0.\label{eq:nonTMG2}
\end{equation}
In (\ref{eq:nonTMG1}) and (\ref{eq:nonTMG2}), one of the equations
can be used in order to determine the constant curvature scalar $R$
in terms of the parameters of the $f\left(R_{\nu}^{\mu}\right)$ theory,
while the other equation is a constraint on the parameters of the
$f\left(R_{\nu}^{\mu}\right)$ theory, as the parameter $\lambda$
in NMG was constrained to take a specific value in terms of $m^{2}$.
Once $R$ is determined, it can be used to determine $m^{2}$ in terms
of the parameters of the $f\left(R_{\nu}^{\mu}\right)$ theory; and
therefore, one obtains the Type-D solutions of the $f\left(R_{\nu}^{\mu}\right)$
theory since, as we noted, $m^{2}$ is the unique parameter appearing
in the Type-D solutions of NMG belonging to these two cases represented
with the equations (\ref{eq:p-R-m2_reln_nonTMG1}) and (\ref{eq:p-R-m2_reln_nonTMG2}).

\section{Applications}

We are now ready to employ the results obtained for the general case
of $f\left(R_{\mu\nu}\right)$ theories in finding the constant scalar
curvature Type-N and Type-D solutions to general quadratic and general
cubic curvature gravity theories and the BINMG theory. We have studied
the solutions of the quadratic curvature gravity as it is the simplest
case for which the Type-N and Type-D solutions can be obtained by
mapping the corresponding solutions of TMG and NMG. Furthermore, the
quadratic curvature gravity is a simple setting for which the explicit
study of the new solutions directly through the field equations is
rather instructive. On the other hand, the solutions for the generic
cubic curvature gravity case naturally provides new solutions to the
cubic curvature extension of NMG which was introduced by using the
holography ideas \cite{Sinha}.%
\footnote{Note that this extension also coincides with small curvature expansion
of BINMG in the third order \cite{BINMG}.%
} Like the cubic curvature case, one can also use the results of the
$f\left(R_{\mu\nu}\right)$ theory in order to study the solutions
of the all higher curvature extensions of NMG based on the holography
ideas given in \cite{Sinha,Paulos}. Finally, BINMG \cite{BINMG}
is an interesting theory either as an infinite order in curvature
extension of NMG which is unitary \cite{Gullu-AllUni3D} and has a
holographic $c$-function matching that of the Einstein's gravity
\cite{Gullu-cfunc}, or on its own right as it appears as a cut-off
independent counterterm to the four dimensional anti-de Sitter space
\cite{Jatkar} and as the first example of a unitary Born-Infeld type
gravity \cite{Deser_Gibbons}. We have provided all Type-N solutions
of BINMG by using the result of \cite{Aliev-PLB}; while the Type-D
solutions that we have found are constrained to the CSI spacetimes.

When one searches for solutions to a given theory in a standard way,
first thing to do is to obtain the field equations which is often
a cumbersome task if the theory involves higher curvature terms. Indeed,
the field equations of the cubic curvature gravity {[}see (\ref{quad_fiel_eq}){]}
and BINMG (see \cite{Gullu-cfunc}) are quite involved. Then, preferably
the field equations are simplified by a suitable choice of ansatz
such as assuming $S_{\mu\nu}$ to be of Type N or Type D as was done
here. Finally, the remaining equations, which still have a nonlinear
complicated form in metric, are needed to be solved. However, by using
the results obtained for generic $f\left(R_{\mu\nu}\right)$ theory,
one bypasses all these complications and obtains the solutions by
mapping the already existing solutions of TMG and NMG via rather simple
relations between the parameters of the theories.

\subsection{Solutions of Quadratic and Cubic Curvature Gravity\label{sec:Solns_quad_curv_on_TMG}}

Since cubic curvature theories include the quadratic ones, we start
with the most general cubic curvature gravity in three dimensions
with the action 
\begin{equation}
I=\int d^{3}x\,\sqrt{-g}\left[\frac{1}{\kappa}\left(R-2\Lambda_{0}\right)+\alpha R^{2}+\beta R_{\nu}^{\mu}R_{\mu}^{\nu}+\gamma_{1}R_{\nu}^{\mu}R_{\mu}^{\rho}R_{\rho}^{\nu}+\gamma_{2}RR_{\nu}^{\mu}R_{\mu}^{\nu}+\gamma_{3}R^{3}\right].\label{eq:Cubic_act}
\end{equation}
In order to use the results of the previous section, first we need
to rewrite (\ref{eq:Cubic_act}) in terms of $\left(R,A,B\right)$
as 
\begin{align}
I=\int d^{3}x\,\sqrt{-g} & \biggl[\frac{1}{\kappa}\left(R-2\Lambda_{0}\right)+\left(\alpha+\frac{\beta}{3}\right)R^{2}+\beta S_{\nu}^{\mu}S_{\mu}^{\nu}\nonumber \\
 & +\gamma_{1}S_{\nu}^{\mu}S_{\mu}^{\rho}S_{\rho}^{\nu}+\left(\gamma_{1}+\gamma_{2}\right)RS_{\nu}^{\mu}S_{\mu}^{\nu}+\left(\frac{\gamma_{1}}{9}+\frac{\gamma_{2}}{3}+\gamma_{3}\right)R^{3}\biggr].\label{eq:Cubic_act_in_Smn}
\end{align}
Type-N solutions of the cubic curvature gravity can be found by solving
the set (\ref{eq:R_eqn_TypeN}) and (\ref{eq:mu_eqn_TypeN}) where
the terms $F$, $F_{R}$ and $F_{A}$ for the cubic curvature gravity
with the Type-N ansatz have the forms
\begin{equation}
F=\frac{1}{\kappa}\left(R-2\Lambda_{0}\right)+\left(\alpha+\frac{\beta}{3}\right)R^{2}+\left(\frac{\gamma_{1}}{9}+\frac{\gamma_{2}}{3}+\gamma_{3}\right)R^{3},
\end{equation}
\begin{equation}
F_{R}=\frac{1}{\kappa}+2\left(\alpha+\frac{\beta}{3}\right)R+\left(\frac{\gamma_{1}}{3}+\gamma_{2}+3\gamma_{3}\right)R^{2},\qquad F_{A}=\beta+\left(\gamma_{1}+\gamma_{2}\right)R.\label{eq:FR_FA_cubic}
\end{equation}
After employing these in (\ref{eq:R_eqn_TypeN}) and (\ref{eq:mu_eqn_TypeN}),
one can find the following relation between the parameters of the
cubic theory and $\mu^{2}$, $\Lambda$;
\begin{equation}
\mu^{2}=-\left[\beta+6\Lambda\left(\gamma_{1}+\gamma_{2}\right)\right]^{-1}\left[\frac{1}{\kappa}+\Lambda\left(12\alpha+5\beta\right)+6\Lambda^{2}\left(3\gamma_{1}+7\gamma_{2}+18\gamma_{3}\right)\right],\label{eq:TypeN_m2_cubic}
\end{equation}
provided that $\beta+6\Lambda\left(\gamma_{1}+\gamma_{2}\right)\ne0$,
and $\Lambda$ should satisfy 
\begin{equation}
\frac{\Lambda-\Lambda_{0}}{2\kappa}-\left(3\alpha+\beta\right)\Lambda^{2}-6\left(\gamma_{1}+3\gamma_{2}+9\gamma_{3}\right)\Lambda^{3}=0,\label{eq:TypeN_L_cubic}
\end{equation}
whose solutions are not particularly illuminating to depict. Therefore,
if $\mu^{2}$ and $\Lambda$ of the NMG Type-N solutions \cite{Aliev-PLB},
which also involve the Type-N solutions of TMG, are tuned with the
parameters of the cubic theory according to (\ref{eq:TypeN_m2_cubic})
and (\ref{eq:TypeN_L_cubic}), then these spacetimes also solve the
cubic theory. Furthermore, setting $\gamma_{1}=\gamma_{2}=\gamma_{3}=0$
yields the Type-N field equations of quadratic curvature gravity whose
solutions are given below in (\ref{eq:Type-N_m2}) and (\ref{eq:Type-N_L}).

Now, moving on to the Type-D case, first one needs to calculate $F$,
$F_{R}$, $F_{A}$, $F_{B}$ from (\ref{eq:Cubic_act_in_Smn}) for
the Type-D spacetime ansatz which become
\begin{gather}
F=\frac{1}{\kappa}\left(R-2\Lambda_{0}\right)+\left(\alpha+\frac{\beta}{3}\right)R^{2}+\left(\frac{\gamma_{1}}{9}+\frac{\gamma_{2}}{3}+\gamma_{3}\right)R^{3}+6\left[\beta+\left(\gamma_{1}+\gamma_{2}\right)R\right]p^{2}-6\gamma_{1}p^{3},\nonumber \\
F_{R}=\frac{1}{\kappa}+2\left(\alpha+\frac{\beta}{3}\right)R+6\left(\gamma_{1}+\gamma_{2}\right)p^{2}+\left(\frac{\gamma_{1}}{3}+\gamma_{2}+3\gamma_{3}\right)R^{2},\label{eq:Fs_of_cubic_TypeD}\\
F_{A}=\beta+\left(\gamma_{1}+\gamma_{2}\right)R,\qquad F_{B}=\gamma_{1}.\nonumber 
\end{gather}
Then, using the calculated forms of $F$, $F_{R}$, $F_{A}$, $F_{B}$
in (\ref{eq:fRmn_TypeD_Eqn1_TMG}) and (\ref{eq:fRmn_TypeD_Eqn2_TMG}),
one obtains
\begin{equation}
-\frac{\Lambda-\Lambda_{0}}{\kappa}+2\left(3\alpha+\beta\right)\Lambda^{2}+12\left(\gamma_{1}+3\gamma_{2}+9\gamma_{3}\right)\Lambda^{3}+\left[\beta+18\left(\gamma_{1}+\gamma_{2}\right)\Lambda\right]p^{2}-3\gamma_{1}p^{3}=0,\label{eq:Type-D_Tr_R3}
\end{equation}
and
\begin{equation}
3\left(3\gamma_{1}-2\gamma_{2}\right)p^{2}-2\left[5\beta+6\left(6\gamma_{1}+5\gamma_{2}\right)\Lambda\right]p-\left[\frac{1}{\kappa}+4\left(3\alpha-\beta\right)\Lambda-12\left(3\gamma_{1}+\gamma_{2}-9\gamma_{3}\right)\Lambda^{2}\right]=0.\label{eq:Type-D_Trless_R3}
\end{equation}
If the parameters of TMG ($\mu^{2}$ and $\Lambda$) are tuned according
to these two algebraic relations in terms of the parameters of the
cubic curvature theory, then the Type-D solutions of TMG also solve
the cubic curvature theory. Solving this set of equations after setting
$\gamma_{1}=\gamma_{2}=\gamma_{3}=0$ yields the quadratic curvature
gravity results that are given below in (\ref{eq:Type-D_m2}) and
(\ref{eq:Type-D_L}).

Although the solutions of the quadratic curvature gravity are obtained
by use of the cubic curvature results, it is rather instructive to
rederive these solutions by using the field equations of the quadratic
curvature gravity. Because, one can arrive at the results by using
the classification scheme via the scalar invariants $I$ and $J$
as described in Sec.~\ref{sec:Algebraic-Classification}, and the
masses of linearized excitations around the (anti)-de Sitter spacetime
appear in the formalism in a rather curious way. First of all, let
us discuss the linearized modes of TMG around (A)dS. Note that there
is a single spin-2 excitation in TMG with mass-squared 
\begin{equation}
m_{\text{TMG}}^{2}=\mu^{2}+\Lambda.\label{eq:MassTMG}
\end{equation}
This can be seen from the linearization of (\ref{box_tmg}) as follows:
the right hand side vanishes for any Einstein space and the left hand
side yields 
\begin{equation}
\left(\bar{\square}-\mu^{2}-3\Lambda\right)S_{\mu\nu}^{L}=0,\label{eq:Lin_of_boxTMG}
\end{equation}
where $\bar{\square}$ refers to the background d'Alembertian with
a metric $\bar{g}_{\mu\nu}$. Keeping in mind that in three-dimensional
AdS spacetime, a \emph{massless} spin-2 field satisfies 
\begin{equation}
\left(\bar{\square}-2\Lambda\right)h_{\mu\nu}^{\text{TT}}=0,\label{eq:EoM_for_massless_spin0}
\end{equation}
where $h_{\mu\nu}^{\text{TT}}$ is the transverse-traceless part of
the tensorial excitation $h_{\mu\nu}$. Comparing (\ref{eq:Lin_of_boxTMG})
and (\ref{eq:EoM_for_massless_spin0}), the mass in (\ref{eq:MassTMG})
follows. Even though this heuristic procedure led to the correct mass,
one should always check such a computation with the help of a through
canonical procedure, since $S_{\mu\nu}^{L}$ is not a fundamental
excitation in this theory. Canonical analysis of this theory was carried
out in \cite{Carlip-CQG,Carlip-PLB} which agrees with our heuristic
derivation.

Now, turning to the discussion on the solutions of the quadratic curvature
gravity. For spacetimes of constant scalar curvature, the trace and
the traceless field equations of the quadratic curvature gravity take
the form
\begin{equation}
-\frac{R-6\Lambda_{0}}{\kappa}+\frac{3\alpha+\beta}{3}R^{2}+\beta S_{\mu\nu}S^{\mu\nu}+2K=0,\label{eq:Tr_quadratic_const_R}
\end{equation}
 and 
\begin{equation}
\left(\beta\square+\frac{1}{\kappa}+\frac{6\alpha+\beta}{3}R\right)S_{\mu\nu}=4\beta\left(S_{\mu\rho}S_{\nu}^{\rho}-\frac{1}{3}g_{\mu\nu}S_{\sigma\rho}S^{\sigma\rho}\right),\label{eq:Trless_quadratic_const_R}
\end{equation}
by using (\ref{pure_trace}) and (\ref{box_traceless_ricci_ten}).
Then, if we require that the solution of the quadratic curvature gravity
is also a solution of the cosmological TMG, one gets a quadratic constraint
\begin{equation}
S_{\mu\rho}S_{\nu}^{\rho}-\frac{1}{3}g_{\mu\nu}S_{\sigma\rho}S^{\sigma\rho}=\left(\frac{1}{\kappa\beta}+\mu^{2}+5\Lambda+\frac{12\Lambda\alpha}{\beta}\right)S_{\mu\nu},\label{eq:Quad_constraint_before_use_of_Tr_eqn}
\end{equation}
by using the field equations of TMG which are $R=6\Lambda$ and (\ref{box_tmg})
in (\ref{eq:Trless_quadratic_const_R}). Besides, one can also use
(\ref{eq:Tr_quadratic_const_R}) in order to further reduce (\ref{eq:Quad_constraint_before_use_of_Tr_eqn})
to 
\begin{equation}
\beta S_{\mu\rho}S_{\nu}^{\rho}-\left(\frac{1}{\kappa}+(\mu^{2}+5\Lambda)\beta+12\Lambda\alpha\right)S_{\mu\nu}+\left(\frac{2}{\kappa}\left(\Lambda_{0}-\Lambda\right)+4\left(3\alpha+\beta\right)\Lambda^{2}\right)g_{\mu\nu}=0.\label{quadratic_constraint}
\end{equation}
This will serve as the main equation in classifying the solutions
of quadratic gravity that are also solutions of TMG. The scalar invariants
$I$ and $J$ can be read from (\ref{quadratic_constraint}) as 
\begin{equation}
I=-\frac{3}{\beta}\left(\frac{2}{\kappa}\left(\Lambda_{0}-\Lambda\right)+4\left(3\alpha+\beta\right)\Lambda^{2}\right),\qquad J=\left(\frac{1}{\kappa\beta}+\mu^{2}+5\Lambda+\frac{12\Lambda\alpha}{\beta}\right)I.
\end{equation}
 With these we can rewrite (\ref{quadratic_constraint}) as 
\begin{equation}
S_{\mu\rho}S_{\nu}^{\rho}-\frac{J}{I}\, S_{\mu\nu}-\frac{I}{3}\, g_{\mu\nu}=0.\label{short_quad}
\end{equation}
It is interesting to note that $\frac{J}{I}$ is exactly the square
of the mass difference of the \emph{spin-2} excitations of TMG and
quadratic gravity theories%
\footnote{There is also a spin-0 excitation of general quadratic gravity theory
with mass $m_{s}^{2}=\frac{1}{\kappa\left(8\alpha+3\beta\right)}-4\Lambda\left(\frac{3\alpha+\beta}{8\alpha+3\beta}\right)$
which decouple in the NMG limit \cite{Gullu_Tekin,Gullu-Canonical}.%
}, namely 
\begin{equation}
J=\left(m_{\text{TMG}}^{2}-m_{\text{quadratic}}^{2}\right)I,
\end{equation}
 where $m_{\text{TMG}}^{2}=\mu^{2}+\Lambda$ \cite{Carlip-CQG,Carlip-PLB}
and $m_{\text{quadratic}}^{2}=-\frac{1}{\kappa\beta}-4\Lambda-\frac{12\Lambda\alpha}{\beta}$
\cite{Gullu-Canonical}.

What is achieved up to this point is that: If a given $S_{\mu\nu}$
satisfies the TMG equations and the quadratic constraint (\ref{quadratic_constraint}),
then it also satisfies the general quadratic gravity equations. Therefore,
we can use the solutions of TMG which were compiled and classified
in \cite{Chow-Classify}. As noted in Sec.~\ref{sec:Algebraic-Classification},
the Type-N spacetimes satisfy $I=J=0$. Then, from (\ref{quadratic_constraint}),
one obtains
\begin{equation}
\mu^{2}=-\frac{1}{\kappa\beta}-\Lambda\left(5+\frac{12\alpha}{\beta}\right),\label{eq:Type-N_m2}
\end{equation}
 where the effective cosmological constant reads
\begin{equation}
\Lambda=\frac{1}{4\left(3\alpha+\beta\right)\kappa}\left(1\pm\sqrt{1-8\left(3\alpha+\beta\right)\kappa\Lambda_{0}}\right).\label{eq:Type-N_L}
\end{equation}
These equations relate the parameters of Type-N solutions of TMG and
the Type-N solutions of the general quadratic theory. In terms of
the massive spin-2 excitations of the theories, Type-N solutions satisfy
the interesting property $m_{\text{TMG}}^{2}=m_{\text{quadratic}}^{2}$.
When one sets $8\alpha+3\beta=0$, one gets the NMG result given in
\cite{Gurses-Killing,Aliev-PRL,Aliev-PLB} after identifying $m^{2}\equiv-\frac{1}{\kappa\beta}$
.

On the other hand, for the Type-D spacetimes, by using (\ref{eq:Can_Smn_TypeD})
and (\ref{eq:mu-R-p}) in (\ref{short_quad}), we have two equations
valid for both Type-D cases 
\begin{equation}
(p^{2}-\frac{J}{I}p-\frac{I}{3})g_{\mu\nu}=0,\qquad(p+\frac{J}{I})\xi_{\mu}\xi_{\nu}=0,
\end{equation}
from which it follows that $I^{3}=6J^{2}$. These relations yield
\begin{equation}
\mu^{2}=-\frac{9}{10\kappa\beta}-\frac{27\Lambda}{5}\left(1+\frac{2\alpha}{\beta}\right),\label{eq:Type-D_m2}
\end{equation}
 where the effective cosmological constant reads 
\begin{equation}
\Lambda=\frac{1}{12\left(2\alpha+\beta\right)\left(\alpha+3\beta\right)\kappa}\left(-2\alpha+9\beta\pm5\sqrt{\beta\left(-2\alpha+3\beta-8\left(2\alpha+\beta\right)\left(\alpha+3\beta\right)\kappa\Lambda_{0}\right)}\right).\label{eq:Type-D_L}
\end{equation}
In the NMG limit, these relations yield the corresponding equations
in \cite{Aliev-PRL,Aliev-PRD}. Let us summarize our results for the
quadratic curvature gravity as:\textit{ Type-N and Type-D solutions
of the TMG also solve the general quadratic gravity if the TMG parameter
and the cosmological constant are tuned as}\emph{ (\ref{eq:Type-N_m2},
\ref{eq:Type-N_L}) and (\ref{eq:Type-D_m2}, \ref{eq:Type-D_L}),
respectively.}

\subsection{Solutions of BINMG}

The action of the Born-Infeld extension of NMG (BINMG) is \cite{BINMG}
\begin{equation}
I_{\text{BI-NMG}}=-\frac{4\tilde{m}^{2}}{\kappa^{2}}\int d^{3}x\,\left[\sqrt{-\det\left(g-\frac{1}{\tilde{m}^{2}}G\right)}-\left(1-\frac{\tilde{\lambda}}{2}\right)\sqrt{-\det g}\right],\label{eq:BI-NMG_action}
\end{equation}
where $G$ is the Einstein tensor with components $G_{\mu\nu}\equiv R_{\mu\nu}-\frac{1}{2}g_{\mu\nu}R$.
Note that we have used the tilded versions of the parameters to avoid
possible confusion with the NMG parameters. By using the exact expansion
\begin{equation}
\det A=\frac{1}{6}\left[\left(\text{Tr}A\right)^{3}-3\text{Tr}A\text{Tr}\left(A^{2}\right)+2\text{Tr}\left(A^{3}\right)\right],
\end{equation}
which is valid for $3\times3$ matrices, (\ref{eq:BI-NMG_action})
can be rewritten as \cite{Gullu-Horava}
\begin{equation}
I_{\text{BI-NMG}}=-\frac{4\tilde{m}^{2}}{\kappa^{2}}\int d^{3}x\,\sqrt{-\det g}F\left(R,A,B\right),\label{eq:BI-NMG_trace}
\end{equation}
 where 
\begin{equation}
F\left(R,A,B\right)\equiv\sqrt{\left(1+\frac{R}{6\tilde{m}^{2}}\right)^{3}-\frac{A}{2\tilde{m}^{4}}\left(1+\frac{R}{6\tilde{m}^{2}}\right)-\frac{B}{3\tilde{m}^{6}}}-\left(1-\frac{\tilde{\lambda}}{2}\right).\label{eq:F_defn}
\end{equation}
Now, let us find the constant curvature Type-N and Type-D solutions
in this theory by using the formalism developed above for the generic
$f\left(R_{\nu}^{\mu}\right)$ theory.

\subsubsection{Type-N solutions}

To begin with, for Type-N spacetimes, the functions $F$, $F_{R}$,
$F_{A}$ take the forms
\begin{equation}
F=\left(1+\frac{\Lambda}{\tilde{m}^{2}}\right)^{\frac{3}{2}}-\left(1-\frac{\tilde{\lambda}}{2}\right),\qquad F_{R}=\frac{1}{4\tilde{m}^{2}}\left(1+\frac{\Lambda}{\tilde{m}^{2}}\right)^{\frac{1}{2}},\qquad F_{A}=-\frac{1}{4\tilde{m}^{4}}\left(1+\frac{\Lambda}{\tilde{m}^{2}}\right)^{-\frac{1}{2}},\label{eq:F_functions_TypeN}
\end{equation}
after setting $R=6\Lambda$ and with the requirement $\Lambda>-\tilde{m}^{2}$
which will be the lower bound on the scalar curvature. Before analyzing
the solutions, let us note an observation. If one uses (\ref{eq:F_functions_TypeN})
in (\ref{eq:TypeN_Trless_field_eqn}), the traceless field equation
of BINMG for Type-N spacetimes of constant curvature becomes
\begin{equation}
\left(\Box-\tilde{m}^{2}-3\Lambda\right)S_{\mu\nu}=0,
\end{equation}
which is the traceless field equation of TMG in the quadratic form
for the same type of spacetimes given in (\ref{eq:Trless_TypeN_TMG})
with $\tilde{m}^{2}=\mu^{2}$. Actually, without any calculation,
one can see that the Type-N solution found in \cite{Aliev-PLB}, where
$\xi_{\nu}=\partial_{v}$ is not a null-Killing vector field, is a
solution of BINMG, since the traceless field equations are equivalent
and the trace field equation just determines the value of the scalar
curvature. More explicitly, one can use the results of Sec.~\ref{sec:Type-N_f(Ricci)}:
inserting (\ref{eq:F_functions_TypeN}) in (\ref{eq:R_eqn_TypeN})
and (\ref{eq:mu_eqn_TypeN}), then solving the resulting equations
yields
\begin{equation}
\Lambda=-\tilde{m}^{2}\tilde{\lambda}\left(1-\frac{\tilde{\lambda}}{4}\right),\qquad\tilde{\lambda}<2,
\end{equation}
and
\begin{equation}
\mu^{2}=\tilde{m}^{2},
\end{equation}
which is the expected result in the light of the observation above.
Let us give the Type-N solution of BINMG, inherited from NMG \cite{Aliev-PLB},
corresponding to negative constant curvature as $\Lambda=-\nu^{2}$:
\begin{equation}
ds^{2}=d\rho^{2}+\frac{2}{\nu^{2}-\beta^{2}}dudv+\left(Z\left(u,\rho\right)-\frac{v^{2}}{\nu^{2}-\beta^{2}}\right)du^{2},\label{eq:TypeN_of_NMG}
\end{equation}
where $\beta$ can be either $\beta=\nu\tanh\left(\nu\rho\right)$
or $\beta=\nu\coth\left(\nu\rho\right)$, and $Z\left(u,\rho\right)$
is
\begin{equation}
Z\left(u,\rho\right)=\frac{1}{\sqrt{\nu^{2}-\beta^{2}}}\left(\cosh\left(\tilde{m}\rho\right)F_{1}\left(u\right)+\sinh\left(\tilde{m}\rho\right)F_{2}\left(u\right)+\cosh\left(\nu\rho\right)f_{1}\left(u\right)+\sinh\left(\nu\rho\right)f_{2}\left(u\right)\right).
\end{equation}
Note that, for BINMG, the solutions corresponding to the special limits
$\tilde{m}^{2}=-\Lambda$ and $\tilde{m}^{2}=0$ are not allowed.
As described in \cite{Aliev-PLB}, the metric
\begin{equation}
ds^{2}=d\rho^{2}+2e^{2\nu\rho}dudv+\left(e^{\nu\rho}\cosh\left(\tilde{m}\rho\right)F_{1}\left(u\right)+e^{\nu\rho}\sinh\left(\tilde{m}\rho\right)F_{2}\left(u\right)+e^{2\nu\rho}f_{1}\left(u\right)+f_{2}\left(u\right)\right)du^{2},
\end{equation}
can be obtained from (\ref{eq:TypeN_of_NMG}) by taking a specific
limit for which $\partial_{v}$ is a Killing vector. This metric represents
the AdS pp-wave solution given in \cite{Alishahiha}. As $\tilde{m}^{2}=-\Lambda$
limit is not possible for BINMG, the corresponding logarithmic solutions
(in the Poincaré coordinates) for the AdS pp-wave solution are not
available as discussed in \cite{Alishahiha}.

\subsubsection{Type-D solutions}

First, one needs to calculate the functions $F$, $F_{R}$, $F_{A}$,
$F_{B}$ for Type-D spacetimes which take the forms
\begin{gather}
F=\sqrt{\left(1+\frac{R}{6\tilde{m}^{2}}+\frac{2}{\tilde{m}^{2}}p\right)\left(1+\frac{R}{6\tilde{m}^{2}}-\frac{p}{\tilde{m}^{2}}\right)^{2}}-\left(1-\frac{\tilde{\lambda}}{2}\right),\nonumber \\
F_{R}=\frac{1}{4\tilde{m}^{2}}\left(F+1-\frac{\tilde{\lambda}}{2}\right)^{-1}\left[\left(1+\frac{R}{6\tilde{m}^{2}}\right)^{2}-\frac{p^{2}}{\tilde{m}^{4}}\right],\label{eq:F_functions_TypeD}\\
F_{A}=-\frac{1}{4\tilde{m}^{4}}\left(F+1-\frac{\tilde{\lambda}}{2}\right)^{-1}\left(1+\frac{R}{6\tilde{m}^{2}}\right),\qquad F_{B}=-\frac{1}{6\tilde{m}^{6}}\left(F+1-\frac{\tilde{\lambda}}{2}\right)^{-1},\nonumber 
\end{gather}
with the requirements $R\ne6\left(p-\tilde{m}^{2}\right)$ and $R>-6\left(\tilde{m}^{2}+2p\right)$
which will provide a bound on the scalar curvature. For Type-D spacetimes
of constant scalar curvature, the traceless field equation of BINMG
can be found by using (\ref{eq:F_functions_TypeD}) in (\ref{eq:TypeD_Trless_field_eqn})
as
\begin{equation}
\left(\Box-\tilde{m}^{2}-\frac{R}{2}+3p\right)S_{\mu\nu}=0,
\end{equation}
which is the traceless field equation of TMG in the quadratic form
for the same type of spacetimes given in (\ref{eq:box_tmg_TypeD-})
with $\tilde{m}^{2}=\mu^{2}$. Again, without any calculation, one
can see that Type-D solutions of TMG are also solutions of BINMG with
the condition (\ref{eq:mu-R-p}) which now reads as
\begin{equation}
p=\frac{\tilde{m}^{2}}{9}+\frac{R}{6},\label{eq:Soln_TypeD_Eqn2}
\end{equation}
but with a constant scalar curvature that is a solution of the trace
field equation of BINMG. Putting this observation aside, one can find
the Type-D solutions of BINMG by directly using the results obtained
in Sec.~\ref{sec:Type-D-f(Ricci)}. In order to have Type-D solutions
of BINMG which are also Type-D solutions of TMG, (\ref{eq:fRmn_TypeD_Eqn1_TMG})
and (\ref{eq:fRmn_TypeD_Eqn2_TMG}) are the equations that need to
be satisfied. Using (\ref{eq:F_functions_TypeD}), (\ref{eq:fRmn_TypeD_Eqn1_TMG})
and (\ref{eq:fRmn_TypeD_Eqn2_TMG}) reduce to
\begin{equation}
\left(F+1-\frac{\tilde{\lambda}}{2}\right)^{-1}\left[\left(1+\frac{R}{6\tilde{m}^{2}}\right)^{2}-\frac{p^{2}}{\tilde{m}^{4}}\right]-\left(1-\frac{\tilde{\lambda}}{2}\right)=0,
\end{equation}
\begin{equation}
\left(F+1-\frac{\tilde{\lambda}}{2}\right)^{-1}\left(1+\frac{R}{6\tilde{m}^{2}}-\frac{p}{\tilde{m}^{2}}\right)\left(1+\frac{3}{2\tilde{m}^{2}}R-\frac{9}{\tilde{m}^{2}}p\right)=0.
\end{equation}
Solution of the second equation is equivalent to (\ref{eq:Soln_TypeD_Eqn2})
with the requirement that the scalar curvature is bounded as $R>-\frac{22}{9}\tilde{m}^{2}$.
Putting this result in the first equation yields the constant scalar
curvature as 
\begin{equation}
R=\frac{9}{16}\tilde{m}^{2}\left[\left(\tilde{\lambda}^{2}-4\tilde{\lambda}-\frac{52}{27}\right)\pm\left(\tilde{\lambda}-2\right)\sqrt{\left(\tilde{\lambda}-\frac{2}{9}\right)\left(\tilde{\lambda}-\frac{34}{9}\right)}\right],\qquad\tilde{\lambda}<\frac{2}{9}.
\end{equation}
The Type-D solutions of TMG is parametrized with $\mu$ and $R$.
Hence, we need to write $\mu$ in terms of the parameters of BINMG
which can be achieved by using (\ref{eq:Soln_TypeD_Eqn2}) in (\ref{eq:mu-R-p}),
and one gets
\begin{equation}
\mu^{2}=\tilde{m}^{2}.
\end{equation}
By using the Type-D solutions of TMG given in \cite{Chow-Classify},
the Type-D solution of BINMG with a timelike Killing vector and a
negative constant scalar curvature $R\equiv-6\nu^{2}$ can be given
as
\begin{equation}
ds^{2}=-\left(dt+\frac{6\tilde{m}}{\tilde{m}^{2}+27\nu^{2}}\cosh\theta d\phi\right)^{2}+\frac{9}{\tilde{m}^{2}+27\nu^{2}}\left(d\theta^{2}+\sinh^{2}\theta d\phi^{2}\right),
\end{equation}
while the Type-D solution of BINMG with a spacelike Killing vector
and a negative constant scalar curvature reads
\begin{equation}
ds^{2}=\frac{9}{\tilde{m}^{2}+27\nu^{2}}\left(-\cosh^{2}\rho d\tau^{2}+d\rho^{2}\right)+\left(dy+\frac{6\tilde{m}}{\tilde{m}^{2}+27\nu^{2}}\sinh\rho d\tau\right)^{2}.
\end{equation}

Now, let us discuss the Type-D solutions of BINMG which are also Type-D
solutions of NMG but not TMG. In order to have such solutions, the
set of equations that need to be satisfied is either (\ref{eq:nonTMG1})
or (\ref{eq:nonTMG2}). In the case of the NMG solution corresponding
to the set (\ref{eq:nonTMG1}), $p=-\frac{R}{3}$ should be satisfied.
Using this condition and (\ref{eq:F_functions_TypeD}), the set (\ref{eq:nonTMG1})
reduces to
\begin{equation}
\tilde{\lambda}=2-2\sqrt{\left(1-\frac{R}{2\tilde{m}^{2}}\right)\left(1+\frac{R}{2\tilde{m}^{2}}\right)^{2}},
\end{equation}
\begin{equation}
\left(F+1-\frac{\tilde{\lambda}}{2}\right)^{-1}\left(1+\frac{R}{2\tilde{m}^{2}}\right)\left(1-\frac{3}{2\tilde{m}^{2}}R\right)=0,
\end{equation}
which has the solution
\begin{equation}
R=\frac{2}{3}\tilde{m}^{2},\qquad\tilde{\lambda}=2-\frac{8\sqrt{6}}{9}.
\end{equation}
The solutions given in \cite{Aliev-PRD} are parametrized with $m$
which is related to $\tilde{m}$ as $m^{2}=\frac{5}{6}\tilde{m}^{2}$.
Then, with the solutions in \cite{Aliev-PRD}, the following two metric
are the solution of BINMG;
\begin{equation}
ds^{2}=-d\tau^{2}+e^{2/\sqrt{3}\tilde{m}\tau}dx^{2}+e^{-2/\sqrt{3}\tilde{m}\tau}dy^{2},
\end{equation}
\begin{equation}
ds^{2}=\cos\left(2/\sqrt{3}\tilde{m}x\right)\left(-dt^{2}+dy^{2}\right)+dx^{2}+2\sin\left(2/\sqrt{3}\tilde{m}x\right)dtdy.
\end{equation}
On the other hand, in the case of the NMG solution corresponding to
the set (\ref{eq:nonTMG2}), the relation between $p$ and $R$ that
should be satisfied is $p=\frac{R}{6}$ for which (\ref{eq:nonTMG2})
reduces to 
\begin{equation}
\tilde{\lambda}=2-2\sqrt{1+\frac{R}{2\tilde{m}^{2}}},\qquad\left(1+\frac{R}{2\tilde{m}^{2}}\right)^{-\frac{1}{2}}=0,
\end{equation}
where the second equation does not have a solution. Therefore, just
like TMG \cite{Aliev_Nutku}, BINMG does not have a constant scalar
curvature Type-D solution with a hypersurface orthogonal Killing vector.

\section{Conclusions}

We have shown that constant scalar curvature Type-N and Type-D solutions
of topologically massive gravity and new massive gravity solve also
the equations of the generic higher curvature gravity built on the
contractions of the Ricci tensor in 2+1 dimensions. Our construction
is based on inheriting the previously studied solutions of the topologically
massive gravity and the new massive gravity. The crux of the argument
presented here is to reduce the highly complicated higher derivative
equations of the $f\left(R_{\mu\nu}\right)$ theory to a nonlinear
wave-like equation in the traceless-Ricci tensor accompanied with
a constant trace equation, and to implement the defining conditions
of the Type-N and Type-D spacetimes along with the condition of the
constancy of the scalar curvature. Save for the actions which include
the contractions of the derivatives of the Ricci tensor, all the three-dimensional
gravity theories that are based on the Ricci tensor are covered in
this work. As explicit examples, we have given the solutions of the
Born-Infeld extensions of the new massive gravity. Note that with
our approach one can also find solutions of the generic $f\left(R_{\mu\nu}\right)$
theory that fall into the other types such as Type-III and Type-I
under the condition that the scalar curvature is constant. In this
work, we have focused on the Type-N and Type-D solutions of the $f\left(R_{\mu\nu}\right)$
theory, since the corresponding solutions of TMG and NMG are well
studied. But, non-constant scalar curvature solutions can be found
by using the techniques developed in \cite{Gurses-Killing}.

\section{\label{ackno} Acknowledgments}

We thank \.{I}.~Güllü for collaboration at the early stages of this
work and A.~Aliev for useful discussions. M.~G. is partially supported
by the Scientific and Technological Research Council of Turkey (TÜB\.{I}TAK)
and Turkish Academy of Sciences (TÜBA). The work of T.~Ç.~\c{S}.
and B.~T. is supported by the TÜB\.{I}TAK Grant No.~110T339. Some
of the calculations in this paper were checked with the help of the
computer package Cadabra \cite{Cadabra1,Cadabra2}.

\appendix

\section{Some Relevant Variations\label{sec:Variations}}

Variations of the three cubic curvature terms are 
\begin{equation}
\delta\left(R_{\nu}^{\mu}R_{\mu}^{\rho}R_{\rho}^{\nu}\right)=3\left[R_{\mu}^{\rho}R_{\rho\alpha}R_{\nu}^{\alpha}+\frac{1}{2}\left(g_{\mu\nu}R^{\beta\rho}R_{\rho}^{\alpha}\nabla_{\beta}\nabla_{\alpha}+R_{\nu}^{\rho}R_{\mu\rho}\Box-2R_{\nu}^{\rho}R_{\rho}^{\alpha}\nabla_{\mu}\nabla_{\alpha}\right)\right]\delta g^{\mu\nu},
\end{equation}
 
\begin{align}
\delta\left(RR_{\nu}^{\mu}R_{\mu}^{\nu}\right)= & R\left[\left(g_{\mu\nu}R^{\alpha\beta}\nabla_{\beta}\nabla_{\alpha}+R_{\mu\nu}\Box-2R_{\nu}^{\alpha}\nabla_{\mu}\nabla_{\alpha}\right)+2R_{\nu}^{\rho}R_{\mu\rho}\right]\delta g^{\mu\nu}\nonumber \\
 & +R_{\beta}^{\alpha}R_{\alpha}^{\beta}\left[\left(g_{\mu\nu}\Box-\nabla_{\mu}\nabla_{\nu}\right)+R_{\mu\nu}\right]\delta g^{\mu\nu},
\end{align}
 
\begin{equation}
\delta\left(R^{3}\right)=3R^{2}\left[\left(g_{\mu\nu}\Box-\nabla_{\mu}\nabla_{\nu}\right)+R_{\mu\nu}\right]\delta g^{\mu\nu}.
\end{equation}

One can calculate $\delta S_{\alpha\beta}$ by using \textbf{
\begin{equation}
\delta R_{\alpha\beta}=\frac{1}{2}\left(g_{\mu\nu}\nabla_{\alpha}\nabla_{\beta}+g_{\mu\alpha}g_{\beta\nu}\Box-g_{\beta\nu}\nabla_{\mu}\nabla_{\alpha}-g_{\alpha\nu}\nabla_{\mu}\nabla_{\beta}\right)\delta g^{\mu\nu},
\end{equation}
}
\begin{equation}
\delta R=\left[R_{\mu\nu}+\left(g_{\mu\nu}\square-\nabla_{\mu}\nabla_{\nu}\right)\right]\delta g^{\mu\nu},
\end{equation}
as 
\begin{align}
\delta S_{\alpha\beta}= & \frac{1}{2}\left[\left(g_{\mu\nu}\nabla_{\alpha}\nabla_{\beta}+g_{\mu\alpha}g_{\beta\nu}\Box-g_{\beta\nu}\nabla_{\mu}\nabla_{\alpha}-g_{\alpha\nu}\nabla_{\mu}\nabla_{\beta}\right)-\frac{2}{3}g_{\alpha\beta}\left(g_{\mu\nu}\square-\nabla_{\mu}\nabla_{\nu}\right)\right]\delta g^{\mu\nu}\nonumber \\
 & +\frac{1}{3}\left[\left(g_{\mu\alpha}g_{\nu\beta}-\frac{1}{3}g_{\alpha\beta}g_{\mu\nu}\right)R-g_{\alpha\beta}S_{\mu\nu}\right]\delta g^{\mu\nu}.
\end{align}
With this result, $\delta A\equiv\delta\left(S_{\beta}^{\alpha}S_{\alpha}^{\beta}\right)$
and $\delta B\equiv\delta\left(S_{\rho}^{\alpha}S_{\alpha}^{\beta}S_{\beta}^{\rho}\right)$
become
\begin{equation}
\delta A=\left[2\left(S_{\mu}^{\alpha}S_{\alpha\nu}+\frac{1}{3}RS_{\mu\nu}\right)+\left(g_{\mu\nu}S^{\alpha\beta}\nabla_{\alpha}\nabla_{\beta}+S_{\mu\nu}\Box-2S_{\nu}^{\alpha}\nabla_{\mu}\nabla_{\alpha}\right)\right]\delta g^{\mu\nu},
\end{equation}
\begin{align}
\delta B= & \left[\frac{3}{2}\left(g_{\mu\nu}S_{\rho}^{\alpha}S^{\beta\rho}\nabla_{\alpha}\nabla_{\beta}+S_{\mu\rho}S_{\nu}^{\rho}\Box-2S_{\rho}^{\alpha}S_{\nu}^{\rho}\nabla_{\mu}\nabla_{\alpha}\right)-S_{\rho}^{\alpha}S_{\alpha}^{\rho}\left(g_{\mu\nu}\square-\nabla_{\mu}\nabla_{\nu}\right)\right]\delta g^{\mu\nu}\nonumber \\
 & +\left[3S_{\mu}^{\rho}S_{\rho}^{\sigma}S_{\nu\sigma}-S_{\rho}^{\alpha}S_{\alpha}^{\rho}S_{\mu\nu}+\left(S_{\mu\rho}S_{\nu}^{\rho}-\frac{1}{3}g_{\mu\nu}S_{\rho}^{\alpha}S_{\alpha}^{\rho}\right)R\right]\delta g^{\mu\nu}.
\end{align}

\section{Field Equations of Cubic Curvature Gravity }

The action (\ref{eq:Cubic_act}) yields the source-free field equations
with the help of the variations above 
\begin{align}
\frac{1}{\kappa}\left(R_{\mu\nu}-\frac{1}{2}g_{\mu\nu}R+\Lambda_{0}g_{\mu\nu}\right)+2\alpha R\left(R_{\mu\nu}-\frac{1}{4}g_{\mu\nu}R\right)+\left(2\alpha+\beta\right)\left(g_{\mu\nu}\square-\nabla_{\mu}\nabla_{\nu}\right)R\label{quad_fiel_eq}\\
+\beta\square\left(R_{\mu\nu}-\frac{1}{2}g_{\mu\nu}R\right)+2\beta\left(R_{\mu\sigma\nu\rho}-\frac{1}{4}g_{\mu\nu}R_{\sigma\rho}\right)R^{\sigma\rho}+K_{\mu\nu} & =0,\nonumber 
\end{align}
 where $\square=\nabla_{\alpha}\nabla^{\alpha}$. Field equation for
the quadratic curvature part is given in \cite{Deser_Tekin} and the
contribution from the cubic curvature part reads 
\begin{align}
K_{\mu\nu}= & \gamma_{1}\left[\frac{3}{2}g_{\mu\nu}\nabla_{\alpha}\nabla_{\beta}\left(R^{\beta\rho}R_{\rho}^{\alpha}\right)+\frac{3}{2}\square\left(R_{\nu}^{\rho}R_{\mu\rho}\right)-3\nabla_{\alpha}\nabla_{(\mu}\left(R_{\nu)}^{\rho}R_{\rho}^{\alpha}\right)+3R_{\mu}^{\rho}R_{\rho\alpha}R_{\nu}^{\alpha}-\frac{1}{2}g_{\mu\nu}R_{\beta}^{\alpha}R_{\alpha}^{\rho}R_{\rho}^{\beta}\right]\nonumber \\
 & +\gamma_{2}\biggl[g_{\mu\nu}\nabla_{\alpha}\nabla_{\beta}\left(RR^{\alpha\beta}\right)+\square\left(RR_{\mu\nu}\right)-2\nabla_{\alpha}\nabla_{(\mu}\left(R_{\nu)}^{\alpha}R\right)+\left(g_{\mu\nu}\square-\nabla_{\mu}\nabla_{\nu}\right)R_{\alpha\beta}^{2}\nonumber \\
 & \phantom{+\gamma_{2}}+2RR_{\nu}^{\rho}R_{\mu\rho}+R_{\mu\nu}R_{\beta}^{\alpha}R_{\alpha}^{\beta}-\frac{1}{2}g_{\mu\nu}RR_{\beta}^{\alpha}R_{\alpha}^{\beta}\biggr]\\
 & +\gamma_{3}\left[3\left(g_{\mu\nu}\Box-\nabla_{\mu}\nabla_{\nu}\right)R^{2}+3R^{2}R_{\mu\nu}-\frac{1}{2}g_{\mu\nu}R^{3}\right].\nonumber 
\end{align}
It is quite useful to recast them into a pure trace and a traceless
part as 
\begin{equation}
\left(8\alpha+3\beta\right)\square R-\frac{R-6\Lambda_{0}}{\kappa}+\frac{3\alpha+\beta}{3}R^{2}+\beta S_{\mu\nu}S^{\mu\nu}+2K=0,\label{pure_trace}
\end{equation}
 and 
\begin{align}
\left(\beta\square+\frac{1}{\kappa}+\frac{6\alpha+\beta}{3}R\right)S_{\mu\nu}= & 4\beta\left(S_{\mu\rho}S_{\nu}^{\rho}-\frac{1}{3}g_{\mu\nu}S_{\sigma\rho}S^{\sigma\rho}\right)+\left(2\alpha+\beta\right)\left(\nabla_{\mu}\nabla_{\nu}-\frac{1}{3}g_{\mu\nu}\square\right)R\label{box_traceless_ricci_ten}\\
 & -\left(K_{\mu\nu}-\frac{1}{3}g_{\mu\nu}K\right),\nonumber 
\end{align}
where $K\equiv g^{\mu\nu}K_{\mu\nu}$. In deriving the quadratic curvature
contribution to these equations which has a Riemann tensor in it (\ref{quad_fiel_eq}),
one makes use of the relation between the Riemann tensor, the traceless
Ricci tensor and the scalar curvature in three dimensions; 
\begin{equation}
R_{\mu\nu\rho\sigma}=g_{\mu\rho}S_{\nu\sigma}+g_{\nu\sigma}S_{\mu\rho}-g_{\nu\rho}S_{\mu\sigma}-g_{\mu\sigma}S_{\nu\rho}+\frac{1}{6}\left(g_{\mu\rho}g_{\nu\sigma}-g_{\mu\sigma}g_{\nu\rho}\right)R.\label{riemann}
\end{equation}
The trace part, $K$, and the traceless part of $K_{\mu\nu}$ are
given in terms of the traceless Ricci tensor as 
\begin{align}
K= & \left(\frac{3}{2}\gamma_{1}+2\gamma_{2}\right)\square\left(S_{\nu}^{\mu}S_{\mu}^{\nu}\right)+\frac{3}{2}\gamma_{1}\left(\nabla_{\alpha}\nabla_{\beta}+S_{\alpha\beta}\right)\left(S^{\alpha\rho}S_{\rho}^{\beta}\right)\nonumber \\
 & +\left(\gamma_{1}+\gamma_{2}\right)\left(\nabla_{\alpha}\nabla_{\beta}+\frac{3}{2}S_{\alpha\beta}\right)\left(RS^{\alpha\beta}\right)+2\left(\frac{\gamma_{1}}{3}+\gamma_{2}+3\gamma_{3}\right)\left(\square+\frac{R}{4}\right)R^{2},
\end{align}
\begin{align}
K_{\mu\nu}-\frac{1}{3}g_{\mu\nu}K= & \left(\square+\frac{2}{3}R\right)\left[\frac{3}{2}\gamma_{1}\left(S_{\mu\rho}S_{\nu}^{\rho}-\frac{1}{3}g_{\mu\nu}S_{\sigma}^{\rho}S_{\rho}^{\sigma}\right)+\left(\gamma_{1}+\gamma_{2}\right)RS_{\mu\nu}\right]\nonumber \\
 & -\nabla_{\alpha}\left(g_{\beta(\nu}\nabla_{\mu)}-\frac{1}{3}g_{\mu\nu}\nabla_{\beta}\right)\left[3\gamma_{1}S_{\rho}^{\alpha}S^{\rho\beta}+2\left(\gamma_{1}+\gamma_{2}\right)RS^{\alpha\beta}\right]\label{eq:cubic}\\
 & +\left(\frac{1}{3}g_{\mu\nu}\square-\nabla_{\mu}\nabla_{\nu}+S_{\mu\nu}\right)\left[\gamma_{2}S_{\sigma}^{\rho}S_{\rho}^{\sigma}+\left(\frac{\gamma_{1}}{3}+\gamma_{2}+3\gamma_{3}\right)R^{2}\right]\nonumber \\
 & +3\gamma_{1}S_{\sigma}^{\rho}\left(S_{\mu\rho}S_{\nu}^{\sigma}-\frac{1}{3}g_{\mu\nu}S_{\rho}^{\alpha}S_{\alpha}^{\sigma}\right)+2\left(\gamma_{1}+\gamma_{2}\right)R\left(S_{\mu\rho}S_{\nu}^{\rho}-\frac{1}{3}g_{\mu\nu}S_{\sigma}^{\rho}S_{\rho}^{\sigma}\right).\nonumber 
\end{align}


\begin{thebibliography}{References}
\bibitem{BHT-PRL} E.~Bergshoeff, O.~Hohm and P.~K.~Townsend,
``Massive Gravity in Three Dimensions,'' Phys.\ Rev.\ Lett., \textbf{102},
201301 (2009).

\bibitem{BHT-PRD} E. Bergshoeff, O. Hohm and P. K. Townsend, ``More
on Massive 3D Gravity,'' Phys.\ Rev.\ D \textbf{79}, 24042 (2009).

\bibitem{Sinha} A.~Sinha, {}``On the new massive gravity and AdS/CFT,''
JHEP \textbf{1006}, 061 (2010).

\bibitem{BINMG} I.~Gullu, T.~Cagri Sisman, B.~Tekin, {}``Born-Infeld
extension of new massive gravity,'' Class.\ Quant.\ Grav.\ \textbf{27},
162001 (2010).

\bibitem{Gullu-AllUni3D} I.~Gullu, T.~C.~Sisman, B.~Tekin, {}``All
Bulk and Boundary Unitary Cubic Curvature Theories in Three Dimensions,''
Phys.\ Rev.\ D\textbf{ 83}, 024033 (2011).

\bibitem{Clement-Warped} G.~Clement, ``Warped AdS(3) black holes
in new massive gravity,'' Class.\ Quant.\ Grav.\ \textbf{26}, 105015
(2009).

\bibitem{Giribet} E.~Ayon-Beato, G.~Giribet and M.~Hassaine, {}``Bending
AdS Waves with New Massive Gravity,'' JHEP \textbf{0905}, 029 (2009).

\bibitem{Clement-Null_Killing} G.~Clement, ``Black holes with a
null Killing vector in new massive gravity in three dimensions,''
Class.\ Quant.\ Grav.\ \textbf{26}, 165002 (2009).

\bibitem{Gurses-Killing} M.~Gurses, ``Killing Vector Fields in Three
Dimensions: A Method to Solve Massive Gravity Field Equations,'' Class.\ Quant.\ Grav.\ \textbf{27},
205018 (2010).

\bibitem{Aliev-PRL} H.~Ahmedov and A.~N.~Aliev, {}``Exact Solutions
in D-3 New Massive Gravity,'' Phys.\ Rev.\ Lett.\ \textbf{106},
021301 (2011).

\bibitem{Aliev-PLB} H.~Ahmedov and A.~N.~Aliev, {}``The General
Type N Solution of New Massive Gravity,'' Phys.\ Lett.\ B \textbf{694},
143 (2010).

\bibitem{Aliev-PRD} H.~Ahmedov and A.~N.~Aliev, {}``Type D Solutions
of 3D New Massive Gravity,'' Phys.\ Rev.\ D \textbf{83}, 084032
(2011).

\bibitem{Bakas} I.~Bakas, C.~Sourdis, {}``Homogeneous vacua of
(generalized) new massive gravity,'' Class.\ Quant.\ Grav.\ \textbf{28},
015012 (2011).

\bibitem{DJT-PRL} S.~Deser, R.~Jackiw and S.~Templeton, {}``Three-Dimensional
Massive Gauge Theories,'' Phys.\ Rev.\ Lett.\ \textbf{48}, 975
(1982).

\bibitem{DJT-Annals} S.~Deser, R.~Jackiw and S.~Templeton, {}``Topologically
Massive Gauge Theories,'' Annals Phys.\ \textbf{140}, 372 (1982)
{[}Erratum-ibid.\ \textbf{185}, 406 (1988){]} {[}Annals Phys.\ \textbf{185},
406 (1988){]} {[}Annals Phys.\ \textbf{281}, 409 (2000){]}.

\bibitem{Gullu_Gurses} I.~Gullu, M.~Gurses, T.~C.~Sisman and
B.~Tekin, ``AdS Waves as Exact Solutions to Quadratic Gravity,''
Phys.\ Rev.\ D \textbf{83}, 084015 (2011).

\bibitem{Pravda} T.~Malek and V.~Pravda, ``Type III and N solutions
to quadratic gravity,'' Phys.\ Rev.\ D \textbf{84}, 024047 (2011).

\bibitem{Nam} S.~Nam, J.~-D.~Park, S.~-H.~Yi, {}``AdS Black
Hole Solutions in the Extended New Massive Gravity,'' JHEP \textbf{1007},
058 (2010).

\bibitem{Ghodsi-BH} A.~Ghodsi, D.~M.~Yekta, {}``Black Holes in
Born-Infeld Extended New Massive Gravity,'' Phys.\ Rev.\ \textbf{D83},
104004 (2011).

\bibitem{Ghodsi-AdS-like} A.~Ghodsi and D.~M.~Yekta, ``On asymptotically
AdS-like solutions of three dimensional massive gravity,'' arXiv:1112.5402
{[}hep-th{]}.

\bibitem{Gullu-cfunc} I.~Gullu, T.~C.~Sisman and B.~Tekin, ``c-functions
in the Born-Infeld extended New Massive Gravity,'' Phys.\ Rev.\ D
\textbf{82}, 024032 (2010).

\bibitem{Aliev-Extended} H.~Ahmedov and A.~N.~Aliev, ``Type N
Spacetimes as Solutions of Extended New Massive Gravity,'' Phys.\ Lett.\ B
\textbf{711}, 117 (2012).

\bibitem{Chow-Classify} D.~D.~K.~Chow, C.~N.~Pope and E.~Sezgin,
``Classification of solutions in topologically massive gravity,''
Class.\ Quant.\ Grav.\ \textbf{27}, 105001 (2010).

\bibitem{Wainwright} J.~Wainwright, ``A Classification Scheme For
Nonrotating Inhomogeneous Cosmologies,'' J.\ Phys.\ A \textbf{12},
2015 (1979).

\bibitem{Hehl} A.~Garcia, F.~W.~Hehl, C.~Heinicke, A.~Macias,
``The Cotton tensor in Riemannian space-times,'' Class.\ Quant.\ Grav.\ \textbf{21},
1099-1118 (2004).

\bibitem{Paulos} M.~F.~Paulos, {}``New massive gravity extended
with an arbitrary number of curvature corrections,'' Phys.\ Rev.\ D
\textbf{82}, 084042 (2010).

\bibitem{Chow-Kundt} D.~D.~K.~Chow, C.~N.~Pope and E.~Sezgin,
``Kundt spacetimes as solutions of topologically massive gravity,''
Class.\ Quant.\ Grav.\textbf{\ 27}, 105002 (2010).

\bibitem{Gurses} M.~Gurses, {}``Perfect fluid sources in 2+1 dimensions,''
Class.\ Quant.\ Grav.\ \textbf{11}, 2585 (1994).

\bibitem{Nutku} Y.~Nutku, ``Exact solutions of topologically massive
gravity with a cosmological constant,'' Class.\ Quant.\ Grav.\ \textbf{10},
2657 (1993).

\bibitem{Jatkar} D.~P.~Jatkar and A.~Sinha, ``New Massive Gravity
and ${\rm AdS}_{4}$ counterterms,'' Phys.\ Rev.\ Lett.\ \textbf{106},
171601 (2011).

\bibitem{Deser_Gibbons} S.~Deser and G.~W.~Gibbons, ``Born-Infeld-Einstein
actions?,'' Class.\ Quant.\ Grav.\ \textbf{15}, L35 (1998).

\bibitem{Gullu_Tekin} I.~Gullu and B.~Tekin, ``Massive Higher Derivative
Gravity in D-dimensional Anti-de Sitter Spacetimes,'' Phys.~Rev.~D
\textbf{80}, 064033 (2009).

\bibitem{Gullu-Canonical} I.~Gullu, T.~C.~Sisman and B.~Tekin,
``Canonical Structure of Higher Derivative Gravity in 3D,'' Phys.\ Rev.\ D
\textbf{81}, 104017 (2010).

\bibitem{Carlip-CQG} S.~Carlip, S.~Deser, A.~Waldron and D.~K.~Wise,
``Cosmological Topologically Massive Gravitons and Photons,'' Class.\ Quant.\ Grav.\ \textbf{26},
075008 (2009).

\bibitem{Carlip-PLB} S.~Carlip, S.~Deser, A.~Waldron and D.~K.~Wise,
``Topologically Massive AdS Gravity,'' Phys.\ Lett.\ B \textbf{666},
272 (2008).

\bibitem{Gullu-Horava} I.~Gullu, T.~C.~Sisman and B.~Tekin, ``Born-Infeld-Horava
gravity,'' Phys.\ Rev.\ D \textbf{81}, 104018 (2010).

\bibitem{Alishahiha} M.~Alishahiha, A.~Naseh and H.~Soltanpanahi,
``On Born-Infeld Gravity in Three Dimensions,'' Phys.\ Rev.\ D \textbf{82},
024042 (2010).

\bibitem{Aliev_Nutku} A.~N.~Aliev and Y.~Nutku, ``A theorem on
topologically massive gravity,'' Class.\ Quant.\ Grav.\ \textbf{13},
L29 (1996).

\bibitem{Cadabra1} K.~Peeters, ``A Field-theory motivated approach
to symbolic computer algebra,'' Comput.\ Phys.\ Commun.\ \textbf{176},
550 (2007).

\bibitem{Cadabra2} K.~Peeters, ``Introducing Cadabra: A Symbolic
computer algebra system for field theory problems,'' arXiv:hep-th/0701238.

\bibitem{Deser_Tekin} S.~Deser, B.~Tekin, ``Energy in generic higher
curvature gravity theories,'' Phys.\ Rev.\ \textbf{D67}, 084009
(2003).\end{thebibliography}
\end{document}